\documentclass[a4paper,12pt,oneoside]{report}

\usepackage[english,greek]{babel}
\usepackage[iso-8859-7]{inputenc}

\usepackage{graphicx}
\usepackage{mathrsfs}
\usepackage{amsfonts}
\usepackage{fancyhdr}
\usepackage{makeidx}
\usepackage{subfigure}
\usepackage{verbatim}
\usepackage{amssymb}
\usepackage{amsmath,bm}
\usepackage{amsthm}
\usepackage{color}
\usepackage{verbatim}

\usepackage{kmath}

\newtheorem{definition}{Ορισμός}[section]
\newtheorem{theorem}{Θεώρημα}[section]
\newtheorem{lemma}{Λήμμα}[section]

\usepackage{algorithm,algpseudocode}

\makeatletter
\renewcommand{\ALG@name}{Αλγόριθμος}
\makeatother

\def\hybrid{\topmargin -30pt    \oddsidemargin 0pt %%%%%%%%%%%%%% Archive-30pt
        \headheight 0pt \headsep 0pt
        \textwidth 6.25in       % A4 paper
        \textheight 9.5in       % A4 paper
        \marginparwidth .875in
        \parskip 5pt plus 1pt   \jot = 1.5ex}

\hybrid

\catcode`\@=11

\def\marginnote#1{}
\newcount\hour
\newcount\minute
\newtoks\amorpm
\hour=\time\divide\hour by60
\minute=\time{\multiply\hour by60 \global\advance\minute by-\hour}
\edef\standardtime{{\ifnum\hour<12 \global\amorpm={am}%
        \else\global\amorpm={pm}\advance\hour by-12 \fi
        \ifnum\hour=0 \hour=12 \fi
        \number\hour:\ifnum\minute<10 0\fi\number\minute\the\amorpm}}
\edef\militarytime{\number\hour:\ifnum\minute<10 0\fi\number\minute}
\def\draftlabel#1{{\@bsphack\if@filesw {\let\thepage\relax
   \xdef\@gtempa{\write\@auxout{\string
      \newlabel{#1}{{\@currentlabel}{\thepage}}}}}\@gtempa
   \if@nobreak \ifvmode\nobreak\fi\fi\fi\@esphack}
        \gdef\@eqnlabel{#1}}
\def\@eqnlabel{}
\def\@vacuum{}
\def\draftmarginnote#1{\marginpar{\raggedright\scriptsize\tt#1}}

\def\draft2{
        \def\@oddfoot{\sl \en preliminary draft \hfil
        \rm\thepage\hfil\sl\today\quad\militarytime}
        \let\@evenfoot\@oddfoot \overfullrule 3pt
        \let\label=\draftlabel
        \let\marginnote=\draftmarginnote
   \def\@eqnnum{(\theequation)\rlap{\kern\marginparsep\tt\@eqnlabel}%
\global\let\@eqnlabel\@vacuum}  }

\def\preprint{\twocolumn\sloppy\flushbottom\parindent 2em
        \leftmargini 2em\leftmarginv .5em\leftmarginvi .5em
        \oddsidemargin -.5in    \evensidemargin -.5in
        \columnsep .4in \footheight 0pt
        \textwidth 10.in        \topmargin  -.4in
        \headheight 12pt \topskip .4in
        \textheight 6.9in \footskip 0pt
        \def\@oddhead{\thepage\hfil\addtocounter{page}{1}\thepage}
        \let\@evenhead\@oddhead \def\@oddfoot{} \def\@evenfoot{} }

\def\numberbysection{\@addtoreset{equation}{section}
        \def\theequation{\thesection.\arabic{equation}}}

\def\underline#1{\relax\ifmmode\@@underline#1\else
        $\@@underline{\hbox{#1}}$\relax\fi}

\def\titlepage{\@restonecolfalse\if@twocolumn\@restonecoltrue\onecolumn
     \else \newpage \fi \thispagestyle{empty}\c@page\z@
        \def\thefootnote{\fnsymbol{footnote}} }

\def\endtitlepage{\if@restonecol\twocolumn \else \newpage \fi
        \def\thefootnote{\arabic{footnote}}
        \setcounter{footnote}{0}}  %\c@footnote\z@ }

\catcode`@=12
\relax

\def\figcap{\section*{Figure Captions\markboth
        {FIGURECAPTIONS}{FIGURECAPTIONS}}\list
        {Figure \arabic{enumi}:\hfill}{\settowidth\labelwidth{Figure
999:}
        \leftmargin\labelwidth
        \advance\leftmargin\labelsep\usecounter{enumi}}}
 \relax
\def\tablecap{\section*{Table Captions\markboth
        {TABLECAPTIONS}{TABLECAPTIONS}}\list
        {Table \arabic{enumi}:\hfill}{\settowidth\labelwidth{Table
999:}
        \leftmargin\labelwidth
        \advance\leftmargin\labelsep\usecounter{enumi}}}
 \relax
\def\reflist{\section*{References\markboth
        {REFLIST}{REFLIST}}\list
        {[\arabic{enumi}]\hfill}{\settowidth\labelwidth{[999]}
        \leftmargin\labelwidth
        \advance\leftmargin\labelsep\usecounter{enumi}}}
 \relax
\makeatletter
\newcounter{pubctr}
\def\publist{\@ifnextchar[{\@publist}{\@@publist}}
\def\@publist[#1]{\list
        {[\arabic{pubctr}]\hfill}{\settowidth\labelwidth{[999]}
        \leftmargin\labelwidth
        \advance\leftmargin\labelsep
        \@nmbrlisttrue\def\@listctr{pubctr}
        \setcounter{pubctr}{#1}\addtocounter{pubctr}{-1}}}
\def\@@publist{\list
        {[\arabic{pubctr}]\hfill}{\settowidth\labelwidth{[999]}
        \leftmargin\labelwidth
        \advance\leftmargin\labelsep
        \@nmbrlisttrue\def\@listctr{pubctr}}}
 \relax
\makeatother

\def\be{\begin{equation}}
\def\ee{\end{equation}}
\def\ba{\begin{eqnarray}}
\def\ea{\end{eqnarray}}

\def\del{\partial}

\def\k{\kappa}
\def\r{\rho}
\def\a{\alpha}
\def\A{\Alpha}
\def\b{\beta}
\def\B{\Beta}
\def\g{\gamma}
\def\G{\Gamma}
\def\d{\delta}
\def\D{\Delta}
\def\e{\epsilon}
\def\E{\Epsilon}
\def\p{\pi}
\def\P{\Pi}
\def\c{\chi}
\def\C{\Chi}
\def\th{\theta}
\def\Th{\Theta}
\def\m{\mu}
\def\n{\nu}
\def\om{\omega}
\def\Om{\Omega}
\def\l{\lambda}
\def\L{\Lambda}
\def\s{\sigma}
\def\S{\Sigma}
\def\vphi{\varphi}

\def\no{\noindent}

\begin{document}

%\draft2
\newcommand{\eqn}[1]{(\ref{#1})}

\linespread{1.3}

\newcommand{\en}{\selectlanguage{english}}
\newcommand{\gr}{\selectlanguage{greek}}
\def\be{\begin{equation}}
\def\ee{\end{equation}}
\def\ba{\begin{eqnarray}}
\def\ea{\end{eqnarray}}
\def\no{\noindent}
\def\del{\partial}

\def\k{\kappa}
\def\r{\rho}
\def\a{\alpha}
\def\A{\Alpha}
\def\b{\beta}
\def\B{\Beta}
\def\g{\gamma}
\def\G{\Gamma}
\def\d{\delta}
\def\D{\Delta}
\def\e{\epsilon}
\def\E{\Epsilon}
\def\p{\pi}
\def\P{\Pi}
\def\c{\chi}
\def\C{\Chi}
\def\th{\theta}
\def\Th{\Theta}
\def\m{\mu}
\def\n{\nu}
\def\om{\omega}
\def\Om{\Omega}
\def\l{\lambda}
\def\L{\Lambda}
\def\s{\sigma}
\def\S{\Sigma}
\def\vphi{\varphi}
\def\half{\frac{1}{2}}
\def \ov {\over}
\def\elF{{\bf F}}
\def\elK{{\bf K}}
\def\elPi{{\bf \Pi}}
\def\elE{{\bf E}}

\def\cA{{\cal A}}
\def\cB{{\cal B}}
\def\cD{{\cal D}}
\def\cF{{\cal F}}
\def\cG{{\cal G}}
\def\cH{{\cal H}}
\def\cL{{\cal L}}
\def\cM{{\cal M}}
\def\cN{{\cal N}}
\def\cO{{\cal O}}
\def\cP{{\cal P}}
\def\cQ{{\cal Q}}
\def\cR{{\cal R}}
\def\cV{{\cal V}}
\def\cY{{\cal Y}}

\hyphenation{δι-πλής  δι-δια-στα-της  δη-μι-ουρ-γί-ας  συ-μπα-γή δια-γρά-μμα-τος
 τα-λά-ντω-σης  συ-μπλη-ρώ-νε-ται   ε-πι-τυγ-χά-νε-ται  τα-χυ-ο-νι-κή  φερ-μι-ο-νι-κός
 συ-σχέ-τι-ση   α-πο-τε-λέ-σμα-τα   κα-τα-σκευ-ής   μι-κρο-σκο-πι-κή   γε-νί-κευ-ση
πα-ρου-σι-ά-ζου-με   ι-δι-ό-τη-τες   πα-ρα-κά-τω  συ-σχε-τί-ζει  κα-τάλ-λη-λους
τε-λε-στές  α-πει-κό-νι-ση   κί-νη-σης  προ-βλέ-πει  λο-γι-σμός  πα-ρου-σι-ά-στη-καν  π
α-ρα-τη-ρού-με  κα-τάλ-λη-λη ε-γκλω-βι-σμός διά-γρα-μμα ε-λεύ-θε-ρη υπερ-βα-ρύ-τη-τα
ε-νε-ργεί-α-κο αντι-κα-τά-στα-ση δια-φέ-ρει ε-γκλω-βι-σμού υ-πό-βα-θρο γε-νι-κευ-μέ-νη
υ-πο-λο-γι-σμέ-νες βα-ρύ-ο-νι-ων συ-μπε-ρι-λα-μβα-νο-μέ-νων δια-χω-ρι-στι-κού
ελα-χι-στο-ποιή-σου-με ελα-χι-στο-ποί-ηση κλα-σι-κή βα-ρυ-ο-νίων βρί-σκου-με
σφαι-ρι-κού συ-νά-ρτη-ση πα-ρα-πά-νω α-ντι-κα-τα-στή-σου-με ι-σο-τρο-πι-κός
προ-σα-να-το-λι-ζό-με-νη α-πο-στά-σεις χρω-μο-δυ-να-μι-κή στοι-χεί-α κβα-ντι-κή Για
υ-πο-λο-γι-σμό ε-πί-σης Κα-ραί-σκο δια-κλά-δω-σης υ-πο-λο-γί-σου-με διε-ξο-δι-κά μια
οι-κο-νο-μι-κή τμή-μα-τος ε-ξε-τά-σου-με δια-κυ-μά-νσε-ων υ-πε-ρί-ωδες προ-σέ-γγι-ζο-ντας
διά-στα-σης αδιά-στα-τες στην κα-νο-νι-κο-ποιή-σου-με πο-λυω-νυ-μι-κή
επα-να-κα-νο-νι-κο-ποιή-σι-μη ενε-ργει-ών σχε-τί-ζο-νται συ-νθή-κη αστα-θές
επα-να-λα-μβά-νο-ντας γνω-στό αρκε-τές έχει θεω-ρία ένα σύ-μμο-ρφη
χα-ρα-κτη-ρι-στι-κό υπά-ρχουν αντι-καθι-στώ-ντας δια-στά-σεις υπερ-φο-ρτία στο
πε-πλε-γμέ-νες πε-πλε-γμέ-νων AdS CFT σύ-μμο-ρφο Coulomb θω-ρα-κι-σμέ-νο
υπε-ριώ-δες προ-σε-γγι-στι-κά δυ-να-μι-κό φο-ρμα-λι-σμό δια-φο-ρε-τι-κές
συ-μπε-ρι-φο-ρά προ-σα-να-το-λι-σμού με-τρι-κής συ-ζευ-γμέ-νοι πα-ρα-μό-ρφω-σης
πα-ρου-σία πα-ρα-μο-ρφω-μέ-νης δια-κυ-μά-νσεις συ-μπε-ρά-σμα-τα αντι-στοι-χίας
απο-κλί-νου-σες εξι-σο-ρρό-πη-ση με-τα-στα-θής αντί-στοι-χη
κα-νο-νι-κο-ποιή-σει συ-νδε-δε-μέ-νο ακτί-νας κό-μβου δια-φο-ρε-τι-κών προ-σω-ρι-νά
μή-κους συ-νο-ψί-ζο-νται Εμμα-νου-ήλ βα-ρυο-νίων χα-ρα-κτη-ρι-στι-κά προ-πτυ-χια-κών
πα-ρα-δεί-γμα-τα γρά-ψου-με δια-τά-ξεις κου-άρκ υ-πο-στή-ρι-ξη δια-τα-ρα-κτι-κές
υπο-λο-γι-σμός ε-ργα-σιών α-ρι-θμού γρα-μμές δια-τα-ρα-κτι-κή}

\author{
	Διπλωματική Εργασία\\
	της \\
	\vspace*{2em}\\
    \Large{Αντωνίας Κορμπά}\\
}

\title{
\begin{figure}[!t]
\begin{center}
\includegraphics[scale=0.5]{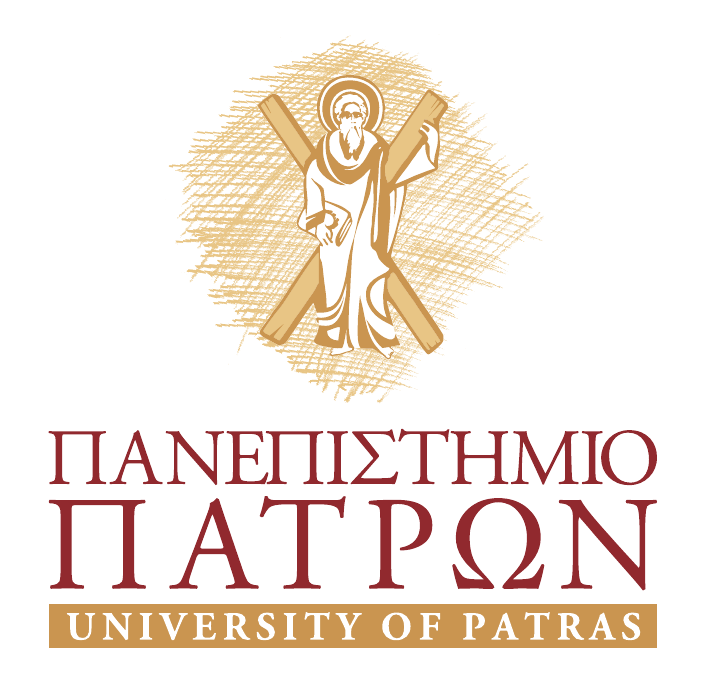}
\end{center}
\begin{center}
\LARGE{\gr \textbf{Μελέτη και Υλοποίηση Αλγορίθμων Κατάταξης σε Διμερή Γραφήματα}}
\end{center}
\label{St-Andreas}
\end{figure}}

\date{\vskip 0.5in
    Τμήμα, Μηχανικών Η/Υ και Πληροφορικής\\
    Πανεπιστήμιο Πατρών\\
    Πάτρα, Ιούλιος 2015
    \pagenumbering{roman}
}

\maketitle

\begin{titlepage}
\begin{center}

\vspace*{2cm}

{\Large Αντωνία Κορμπά}

\vspace{2cm}

{\LARGE{Μελέτη και Υλοποίηση Αλγορίθμων Κατάταξης σε Διμερή Γραφήματα}}

\vspace{2cm}

\vspace{2cm}

\begin{tabular}{ll}
Επιβλέπων :  Καθηγητής Γιάννης Γαροφαλάκης\\
\\
\end{tabular}

\vspace{5cm}

Πάτρα, Ιούλιος 2015
\end{center}
\end{titlepage}

%\pagestyle{headings}
%\renewcommand{\chaptermark}[1]{%
%\markboth{#1}{}}
%\renewcommand{\sectionmark}[1]{%
%\markright{\thesection\ #1}}
%\fancyhf{} % delete current header and footer
%\fancyhead[LE,RO]{\textsl\thepage}
%\fancyhead[LO]{\textsl\rightmark}
%\fancyhead[RE]{\textsl{\leftmark}}
%\renewcommand{\headrulewidth}{0.5pt}
%\renewcommand{\footrulewidth}{0pt}
%\addtolength{\headheight}{0.5pt} % space for the rule
%\fancypagestyle{underline}
%\fancyhead{} % get rid of headers on plain pages
%\renewcommand{\headrulewidth}{0pt} % and the line}

\tableofcontents
\newpage
\newpage

\pagenumbering{arabic}

\gr
\chapter*{Περίληψη}
\addcontentsline{toc}{chapter}{\gr Περίληψη}
Τα τελευταία χρόνια, τα διμερή γραφήματα χρησιμοποιούνται ευρέως σε εφαρμογές ανάκτησης πληροφορίας για να αναπαραστήσουν σχέσεις μεταξύ δύο ομάδων αντικειμένων. O Παγκόσμιος Ιστός μπορεί να προσφέρει μια μεγάλη γκάμα δεδομένων που μπορούν να αναπαρασταθούν από διμερή γραφήματα, όπως είναι ταινίες και κριτικές σε συστήματα προτάσεων, ερωτήματα και σελίδες σε μηχανές αναζήτησης, χρήστες και αναρτήσεις σε μέσα κοινωνικής δικτύωσης. Το μέγεθός και η δυναμική φύση των γραφημάτων αυτών υπαγορεύουν την εύρεση πιο αποδοτικών μεθόδων κατάταξης.

Στην παρούσα διπλωματική εργασία, αρχικά παρουσιάζουμε το βασικό μαθηματικό υπόβαθρο που χρησιμοποιούμε στη συνέχεια και παραθέτουμε τα βασικά στοιχεία της θεωρίας \en Perron-Frobenius \gr για μη αρνητικά μητρώα, καθώς επίσης και της θεωρίας των αλυσίδων  \en Markov\gr. Έπειτα, προτείνουμε έναν νέο αλγόριθμο με όνομα \en BipartiteRank\gr, ο οποίος είναι κατάλληλος για κατάταξη σε διμερή γραφήματα. Ο αλγόριθμος αυτός είναι βασισμένος στο μοντέλο τυχαίου περιπάτου και κληρονομεί τα βασικά μαθηματικά χαρακτηριστικά του \en PageRank\gr. Αυτό που τον διαφοροποιεί, είναι το γεγονός ότι εισάγει ένα άλλο είδος τηλεμεταφοράς που βασίζεται στην \en block \gr δομή του διμερούς γραφήματος για να πετύχει πιο αποδοτική κατάταξη. Τέλος, υποστηρίζουμε την άποψη αυτή με μαθηματικά επιχειρήματα και στη συνέχεια την επιβεβαιώνουμε  και πειραματικά, εκτελώντας μία σειρά από πειράματα σε πραγματικά δεδομένα.

\en
\chapter*{Abstract}
\addcontentsline{toc}{chapter}{\en Abstract}
Recently bipartite graphs have been widely used to represent the relationship two sets of items for information retrieval applications. The Web offers a wide range of data which can be represented by bipartite graphs, such us movies and reviewers in recomender systems, queries and URLs in search engines, users and posts in social networks. The size and the dynamic nature of such graphs generate the need for more efficient ranking methods.

In this thesis, at first we present the fundamental mathematical backround that we use subsequently and we describe the basic principles of the Perron-Frobebius theory for non negative matrices as well as the the basic principles of the Markov chain theory. Then, we propose a novel algorithm named  BipartiteRank, which is suitable to rank scenarios, that can be represented as a bipartite graph. This algorithm is based on the random surfer model and inherits the basic mathematical characteristics of PageRank. What makes it different, is the fact that it introduces an alternative type of teleportation, based on the block structure of the bipartite graph in order to achieve more efficient ranking. Finally, we support this opinion with mathematical arguments and then we confirm it experimentally through a series of tests on real data. \gr

\chapter*{Ευχαριστίες}
Θα ήθελα να ευχαριστήσω όλους όσους συνέβαλαν με οποιονδήποτε τρόπο στην
επιτυχή εκπόνηση αυτής της διπλωματικής εργασίας. 

Καταρχήν, θα πρέπει να ευχαριστήσω καθηγητή κ. Ιωάννη Γαροφαλάκη για την ανάθεση και 
επίβλεψη αυτής της διπλωματικής εργασίας. 

Στη συνέχεια, ευχαριστώ ιδιαίτερα τον διδάκτορα 
Αθανάσιο Νικολακόπουλο για την εξαιρετική
συνεργασία που είχαμε, και ελπίζω να συνεχίσουμε να έχουμε στο μέλλον. Τον
ευχαριστώ θερμά για τις ιδέες που μου προσέφερε καθ'όλη τη διάρκεια εκπόνησης αυτής της
διπλωματικής εργασίας. 

Τέλος, δε μπορώ να μην ευχαριστήσω στην οικογένεια μου που ήταν δίπλα 
μου σε κάθε μου βήμα.

\listoffigures
\listoftables

\newpage

\chapter{Εισαγωγή}

Η κατάταξη αντικειμένων με βάση κάποια κριτήρια είναι ένα πρόβλημα που συναντάμε αρκετά συχνά στην καθημερινότητά μας και ειδικότερα, αποτελεί  αναπόσπαστο κομμάτι ενός συστήματος ανάκτησης πληροφορίας. Στην περίπτωση της αναζήτησης στο διαδίκτυο, λόγω του μεγέθους του Παγκόσμιου ιστού και της ιδιαίτερης φύσης των χρηστών του, ο ρόλος της κατάταξης έχει γίνει τα τελευταία χρόνια ακόμα πιο καθοριστικός και δημιουργήθηκε η ανάγκη για πιο ποιοτικές και αποδοτικές μεθόδους.

Στην παρούσα διπλωματική εργασία, θα βασιστούμε στον πιο γνωστό και ευρέως χρησιμοποιούμενο αλγόριθμο κατάταξης, τον  \en PageRank\gr, προκειμένου να προτείνουμε έναν νέο αλγόριθμο για κατάταξη σε αντικείμενα που σχηματίζουν διμερή γραφήματα. Η ιδέα προήλθε από τον μεταγενέστερο αλγόριθμο κατάταξης, τον \en NCDawareRank\gr και τον τρόπο με τον οποίο εκμεταλλεύεται την ιεραρχική διάρθρωση του χώρου αντικειμένων στον οποίο εφαρμόζεται.

\section{Ανάλυση Υπερσυνδέσμων και Κατάταξη}

Αρχικά, είναι απαραίτητο να διευκρινιστεί τι είναι ανάλυση υπερσυνδέσμων \cite{baeza2011modern} και πώς ένας αλγόριθμος μπορεί να εξάγει την κατάταξη των κορυφών ενός γραφήματος εκμεταλλευόμενος την τοπολογική του δομή. Η ιδέα ξεκίνησε από την παρατήρηση πως μεγάλες συλλογές κειμένων όπως ο Παγκόσμιος Ιστός, επιτρέπουν την ανάπτυξη αλγορίθμων κατάταξης, οι οποίοι λαμβάνουν υπόψη την τοπολογική πληροφορία που παρέχει το γράφημα υπερσυνδέσμων\cite{henzinger2005hyperlink}. Το γράφημα υπερσυνδέσμων του Παγκόσμιου Ιστού, αποτελείται από σελίδες και τα μεταξύ τους \en links\gr. Μία κορυφή αναπαριστά μια σελίδα και μια ακμή ένα \en link \gr. Η πρωταρχική υπόθεση που έγινε, είναι πως ένα εισερχόμενο \en link \gr είναι ένα είδος έγκρισης για μια σελίδα και πιο συγκεκριμένα, όσο περισσότερα εισερχόμενα \en links \gr έχει μια σελίδα τόσο μεγαλύτερο μπορεί να είναι το κύρος της. Πάνω σε αυτή την υπόθεση οι \en S.  Brin \gr και \en L. Page \gr ανέπτυξαν τον αλγόριθμο \en PageRank \gr και εισήγαγαν το \textit{μοντέλο τυχαίου περιηγητή}, ο οποίος ακολουθεί τυχαία τις εξερχόμενες ακμές του γραφήματος του Παγκόσμιου Ιστού περνώντας από άλλες κορυφές συχνά ενώ από άλλες πιο σπάνια. Όπως είναι λογικό, οι κορυφές με πολλές εισερχόμενες ακμές προσπελάζονται συχνά στον τυχαίο περίπατο άρα θεωρείται ότι έχουν μεγαλύτερο κύρος.

Στη φύση υπάρχουν και άλλα γραφήματα όπως αυτό του Παγκόσμιου Ιστού που μπορούν να αντιμετωπιστούν με τον ίδιο τρόπο. Αρκεί κανείς να δει τα δεδομένα από την ανάλογη οπτική γωνία, θεωρώντας πως οι κορυφές είναι αντικείμενα και οι ακμές οποιεσδήποτε αλληλεπιδράσεις μεταξύ των αντικειμένων. Άλλωστε, αλγόριθμοι όπως ο \en PageRank \gr έχουν πάρει τέτοια έκταση που χρησιμοποιούνται πλέον σε μεγάλη ποικιλία προβλημάτων κατάταξης.

\section{Διμερή Γραφήματα}

Ποίο είναι το κίνητρο που μας ώθησε στο να ασχοληθούμε με διμερή γραφήματα$;$ Πόσο συχνά χρησιμοποιούνται σε εφαρμογές και ποια είναι τα βασικότερα χαρακτηριστικά της δομής τους όταν τα συναντάμε στον πραγματικό κόσμο$;$

Τα διμερή γραφήματα χρησιμοποιούνται ευρέως στην αναζήτηση στο διαδίκτυο και σε εφαρμογές ανάκτησης πληροφορίας για να αναπαραστήσουν σχέσεις μεταξύ δύο ομάδων αντικειμένων. O Παγκόσμιος Ιστός μπορεί να προσφέρει μια μεγάλη γκάμα δεδομένων που μπορούν να αναπαρασταθούν από διμερή γραφήματα, όπως είναι ταινίες και κριτικές σε συστήματα προτάσεων, ερωτήματα και σελίδες σε μηχανές αναζήτησης, χρήστες και προϊόντα σε διαδικτυακά καταστήματα, χρήστες και αναρτήσεις σε μέσα κοινωνικής δικτύωσης και λοιπά.
	
\begin{figure}[ht]
  \centering
      \includegraphics[width=0.8\textwidth]{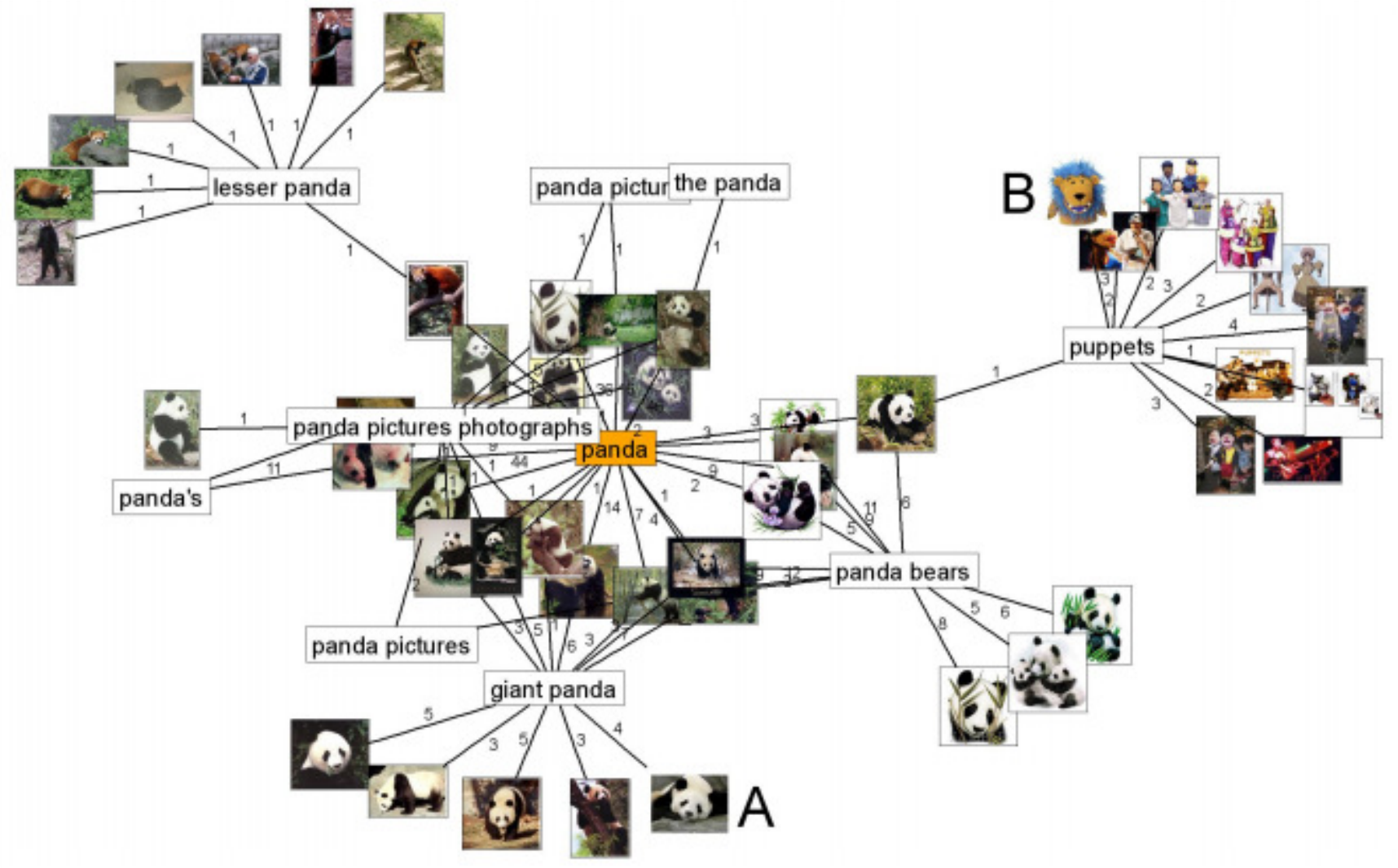}
  \caption{Παράδειγμα διμερούς γραφήματος με ερωτήματα και εικόνες. Η πηγή από την οποία αντλήθηκε η εικόνα είναι η ερευνητική εργασία \cite{craswell2007random}.}
\end{figure} \label{clickGraph}
	
Στα \cite{craswell2007random, deng2009generalized} εφαρμόζεται τυχαίος περίπατος σε διμερή γραφήματα που προέρχονται από μηχανές αναζήτησης. Πιο συγκεκριμένα, οι συγγραφείς του \cite{craswell2007random}, αναφέρουν πως μια μηχανή αναζήτησης μπορεί να καταγράψει ποια από τα αποτελέσματά επιλέχθηκαν από έναν συγκεκριμένο χρήστη για ένα συγκεκριμένο ερώτημα. Για μια δημοφιλή αναζήτησης, η καταγραφή αυτή μπορεί να περιλαμβάνει εκατομμύρια ζεύγη ερωτημάτων-σελίδων ημερησίως, τα οποία σχηματίζουν διμερή γραφήματα. Ένα παράδειγμα απεικονίζεται στο σχήμα $\ref{clickGraph}$. Οι ισχυρισμοί για την χρησιμότητα τέτοιου είδους γραφημάτων σε εφαρμογές αναζήτησης επιβεβαιώνονται και στα \cite{deng2009generalized, li2008learning}. Βέβαια η διαδικασία που ακολουθείται προκειμένου να σχηματιστεί το διμερές γράφημα δεν είναι πάντα ίδια και εξαρτάται πάντα από τη φύση της εφαρμογής και τους εκάστοτε στόχους.

Άλλη μια σημαντική αναφορά στην εφαρμογή τυχαίου περιπάτου σε διμερή γραφήματα παρουσιάζεται στο \cite{cheng2007recommendation}, όπου το ενδιαφέρον στρέφεται στα συστήματα προτάσεων. Σε τέτοιες εφαρμογές οι κορυφές του διμερούς γραφήματος αντιπροσωπεύουν χρήστες και αντικείμενα, ενώ οι ακμές τις αλληλεπιδράσεις μεταξύ τους. Η χρήση του τυχαίου περιπάτου για την κατάταξη των κορυφών με βάση τον κύρος, θεωρείται πως έχει πολύτιμα πλεονεκτήματα αφού λαμβάνει υπόψη τις σχέσεις μεταξύ χρηστών και αντικειμένων καθολικά και χωρίς να αγνοεί χρήσιμη πληροφορία, αντίθετα με άλλες τοπικές μεθόδους.  Το επιχείρημα αυτό ενισχύεται ακόμα περισσότερο και στο \cite{gori2007itemrank}. Επίσης, σύμφωνα με την ερευνητική εργασία \cite{cheng2007recommendation}, εκτός από χρήστες και αντικείμενα υπάρχουν και επιπλέον χαρακτηριστικά που μπορούν να χρησιμοποιηθούν ώστε να βελτιωθεί ακόμα περισσότερο η ποιότητα προτάσεων. Για παράδειγμα, σε προτάσεις προϊόντων τα επιπλέον χαρακτηριστικά μπορεί να είναι το είδος, η μάρκα, η χώρα προέλευσης και λοιπά. Ένας φυσικός τρόπος απεικόνισης όλων αυτών των χαρακτηριστικών σε ένα γράφημα είναι η κατασκευή κ-μερών γραφημάτων. Στα πειράματα που θα πραγματοποιήσουμε στη συνέχεια, θα χρησιμοποιήσουμε μεταξύ άλλων και διμερή γραφήματα που προέρχονται από συστήματα προτάσεων.

Επιπλέον, τα τελευταία χρόνια, ιδιαίτερο ενδιαφέρον παρουσιάζουν τα γραφήματα των οποίων οι κορυφές αντιπροσωπεύουν χρήστες και κοινότητες χρηστών και χρησιμοποιούνται συνήθως στη μελέτη κοινωνικών δικτύων. Μάλιστα, πολλές φορές στη βιβλιογραφία αναφέρονται ως  \en affiliation networks\gr. Η εφαρμογή μεθόδων κατάταξης τέτοιου είδους γραφήματα θα ήταν πολύ χρήσιμη ώστε να εξαχθούν συμπεράσματα όπως, ποιες είναι η πιο δημοφιλείς κοινότητες, ποιοι είναι οι χρήστες με τη μεγαλύτερη συμμετοχή και λοιπά. Οι ερευνητικές εργασίες \cite{newman2002random} και \cite{lattanzi2009affiliation} αναφέρουν κάποια ενδιαφέροντα χαρακτηριστικά της δομής τους όπως είναι οι κατανομές εισόδου των κορυφών χωρίς όμως να κάνουν λόγο για εφαρμογή τυχαίου περιπάτου.

\subsection{Πυκνότητα}

\begin{figure}
\includegraphics[width=\textwidth]{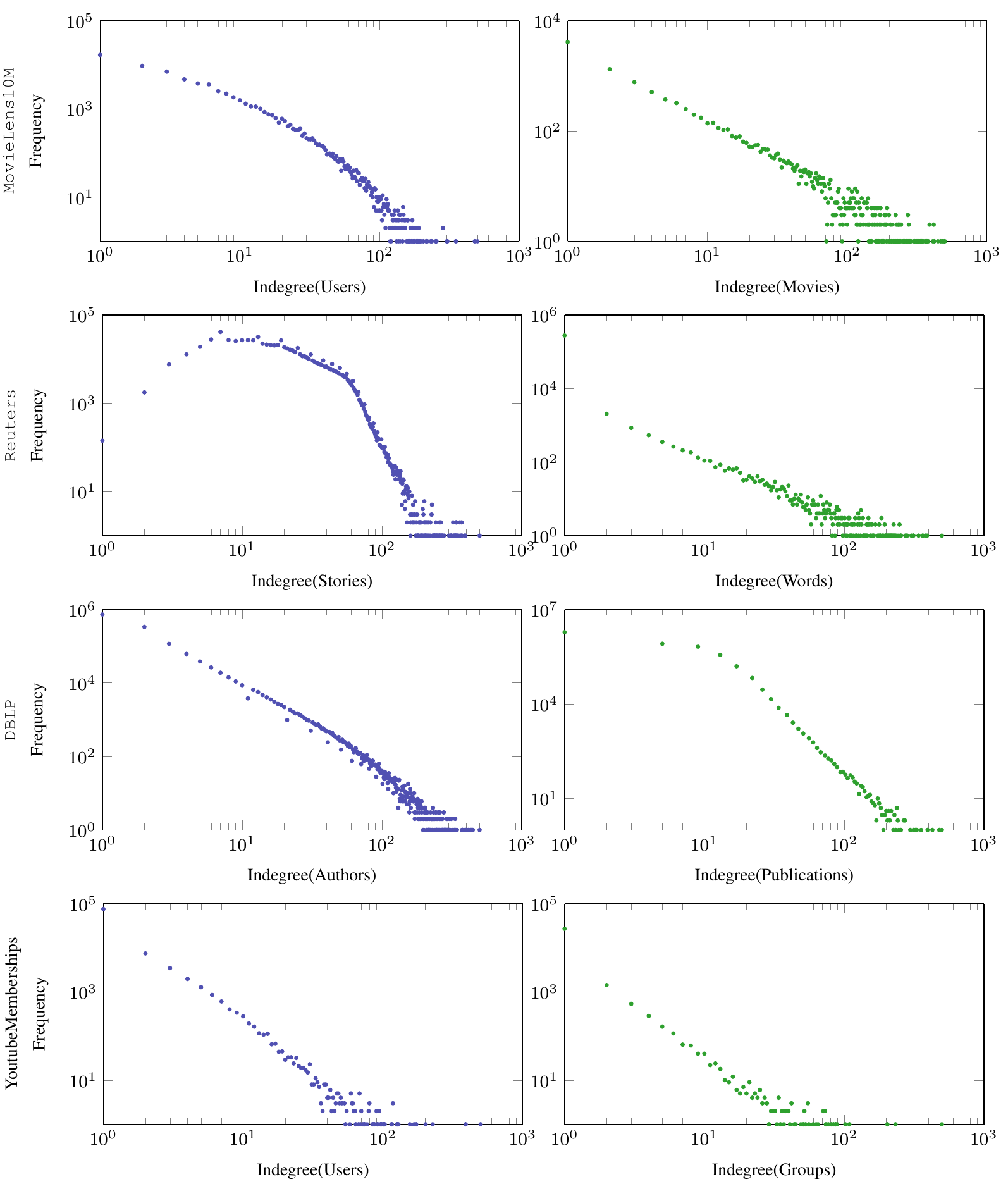}
\caption{Κατανομές εισόδου που προέρχονται από τέσσερα σύνολα δεδομένων.} \label{fig:DegreeDistributions}
\end{figure} 

Σε ένα διμερές γράφημα υπάρχουν δύο κατανομές εισόδου, μία για κάθε σύνολο κορυφών. Σύμφωνα με το \cite{newman2002random} οι κατανομές αυτές σε διμερή γραφήματα του πραγματικού κόσμου δεν είναι τόσο ομοιόμορφες όσο θα μπορούσε κάποιος να υποθέσει. Στην πραγματικότητα είναι ασύμμετρες και σε αρκετές εφαρμογές παρουσιάζονται να ακολουθούν έναν εκθετικό νόμο \cite{lattanzi2009affiliation}. Πιο συγκεκριμένα, αυτό σημαίνει πως ο αριθμός των κορυφών με $i$ γείτονες είναι ανάλογος του $i^{-\gamma}$ για μια σταθερά $\gamma$. Στο σχήμα $\ref{fig:DegreeDistributions}$ μπορούμε να δούμε ενδεικτικά τις κατανομές εισόδου των δεδομένων \en MovieLens10M, Reuteurs, DBLP \gr και \en YoutubeMemberships \gr που αφορούν σε γραφήματα χρηστών-ταινιών, κειμένων-λέξεων, συγγραφέων-δημοσιεύσεων και χρηστών-ομάδων χρηστών αντίστοιχα. Παρατηρούμε πως παρόλο που φύση των δεδομένων είναι διαφορετική, οι κατανομές μοιάζουν αρκετά και έχουν τα χαρακτηριστικά που περιγράψαμε παραπάνω. Συνεπώς, θα πρέπει να σημειώσουμε πως τα γραφήματα στα οποία θα εφαρμόσουμε τη μέθοδο κατάταξης μας είναι ιδιαίτερα αραιά.

\subsection{Μη Συνδεδεμένα Τμήματα} \label{sub:disconnectedcomponents}

Όπως όλα τα γραφήματα του πραγματικού κόσμου, έτσι και τα διμερή γραφήματα μπορεί να είναι μη συνδεδεμένα, δηλαδή να διαθέτουν περισσότερα από ένα συνδεδεμένα τμήματα. Ωστόσο, σύμφωνα με την ερευνητική εργασία \cite{newman2002random}, τα περισσότερα μη συνδεδεμένα γραφήματα που υπάρχουν στην κοινωνία και τη φύση, έχουν ένα μεγάλο συνδεδεμένο τμήμα, γνωστό και ως \en giant component \gr. Τα γραφήματα που δεν έχουν αυτό το χαρακτηριστικό είναι πιο σπάνια. Συνήθως, το μεγαλύτερο συνδεδεμένο τμήμα φέρει πληροφορία που χαρακτηρίζει ολόκληρο το γράφημα \cite{leskovec2005graphs}.

\begin{figure}
	\centering
    \includegraphics[width=1\textwidth]{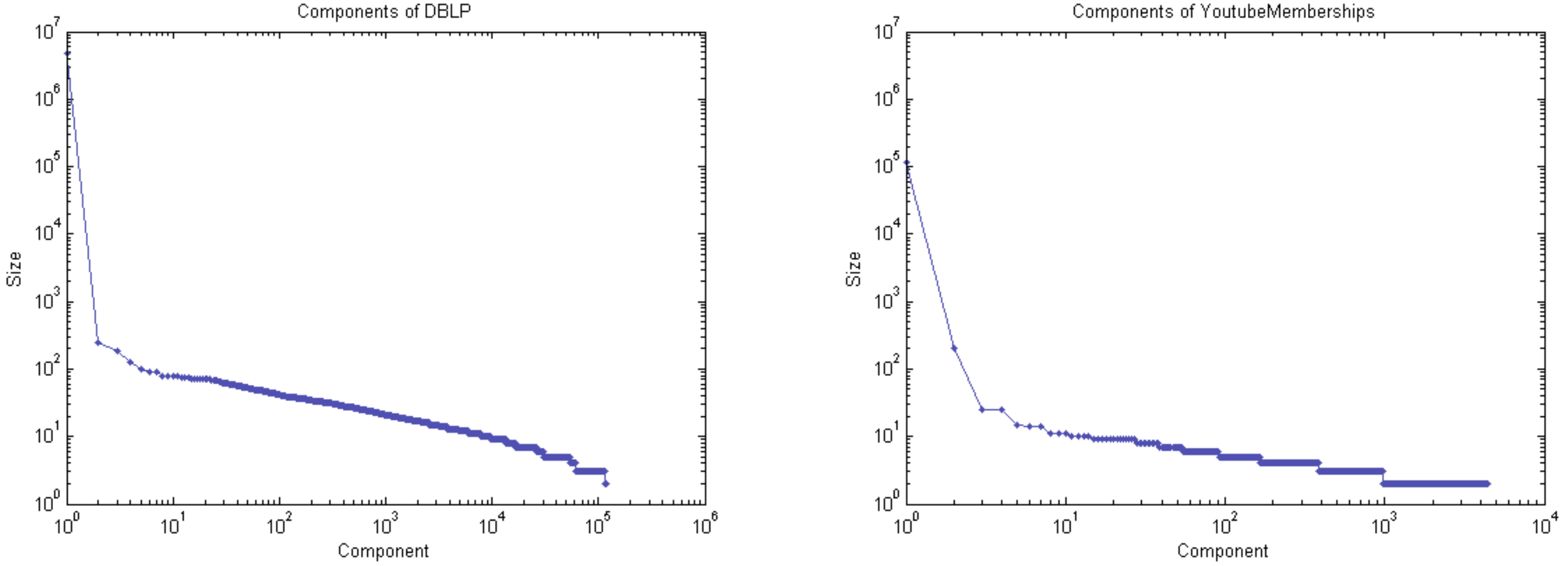}
    \caption{Μεγέθη μη συνδεδεμένων τμημάτων τμημάτων των γραφημάτων \en DBLP \gr και \en YoutubeMemberships\gr.}\label{fig:ComponentSizes}
\end{figure}

Μελετήσαμε τα μεγέθη των μη συνδεδεμένων τμημάτων δύο συνόλων δεδομένων, των \en DBLP \gr και \en YoutubeMemberships\gr. Όπως φαίνεται στο σχήμα $\ref{fig:ComponentSizes}$, έχουν ένα συνδεδεμένο τμήμα αρκετά μεγαλύτερο από τα υπόλοιπα. Πιο συγκεκριμένα, το $88,77 \%$ των κορυφών του \en DBLP \gr ανήκουν στο μεγαλύτερο συνδεδεμένο τμήμα του, ενώ το αντίστοιχό ποσοστό για το \en YoutubeMemberships \gr είναι $91,29 \%$.

\section{Προηγούμενα αποτελέσματα}

Η μέθοδος που θα προτείνουμε στη συνέχεια είναι μια μέθοδος που θα μπορούσε να εφαρμοστεί σε οποιοδήποτε διμερές γράφημα, ανεξάρτητα από το το τι αντιπροσωπεύουν τα σύνολα του. Έτσι, λόγω του γενικού χαρακτήρα του προβλήματος κατάταξης που εξετάζουμε, θεωρούμε πως ο καταλληλότερος αλγόριθμος που θα μπορούσε να εφαρμοστεί σε γενικές περιπτώσεις διμερών γραφημάτων είναι ο αλγόριθμος \en PageRank \gr, αφού χρησιμοποιεί μόνο την τοπολογική δομή του εκάστοτε γραφήματος. Επιπλέον, έχει προταθεί αρκετές φορές η χρήση του για κατάταξη σε πιο συγκεκριμένες εφαρμογές, είτε σε συνδυασμό με άλλες μεθόδους (π.χ.  \en HITS \gr \cite{kleinberg1999authoritative}), \cite{zhou2007co, deng2009generalized}, είτε σε παραλλαγές \cite{hotho2006folkrank,gori2007itemrank}.

Ο αλγόριθμος \en FolkRank \gr \cite{hotho2006folkrank} είναι μία προσαρμοσμένη εκδοχή του \en PageRank\gr. Μπορεί να χρησιμοποιηθεί στην αναζήτηση σε \en folksonomies\gr. Ένα \en folksonomy \gr περιγράφει χρήστες, αντικείμενα, \en tags \gr και την αντιστοίχηση των \en tags \gr σε αντικείμενα με βάση τον χρήστη (βλέπε \cite{hotho2006folkrank} για περισσότερες πληροφορίες). Μπορεί επίσης να αναπαρασταθεί με τη βοήθεια τριμερών γραφημάτων τα οποία παρουσιάζουν κοινά χαρακτηριστικά με τα διμερή γραφήματα. Ο αλγόριθμος \en ItemRank \gr \cite{gori2007itemrank} αποτελεί μια ακόμα ενδιαφέρουσα παραλλαγή, και χρησιμοποιείται σε συστήματα προτάσεων. Ο αλγόριθμος αυτός κατατάσσει προϊόντα με βάση τις αναμενόμενες προτιμήσεις ενός χρήστη, με άλλα λόγια  προτείνει σχετικά αντικείμενα σε εν δυνάμει ενδιαφερόμενους χρήστες. 

\section{Σημειογραφία}

Στη συνέχεια, αναπαριστούμε τα διανύσματα με έντονους πεζούς χαρακτήρες (π.χ. $\boldsymbol{\pi}$), τα μητρώα με έντονους κεφαλαίους χαρακτήρες (π.χ. $\mathbf{P}$). Αναπαριστούμε την  $i^{\text{ή}}$ γραμμή και $j^{\text{ή}}$ στήλη ενός μητρώου $\mathbf{P}$ με $\mathbf{p}^\intercal_{i}$ και $\mathbf{p}_{j}$ αντίστοιχα, και το $ij^{ό}$ στοιχείο του μητρώου $\mathbf{P}$ με $\mathbf{P}_{i,j}$. Επίσης , χρησιμοποιούμε καλλιγραφικά γράμματα για να αναπαραστήσουμε σύνολα (π.χ., $\mathcal{U,V}$). Τέλος, αναπαριστούμε με $[1,n]$ ένα σύνολο ακεραίων $\{1,2,\dots,n\}$.

\section{Οργάνωση}

Στο \textbf{κεφάλαιο 2} παραθέτουμε το μαθηματικό υπόβαθρο της διπλωματικής αυτής, ώστε να γίνουν κατανοητά τα μαθηματικά των μεθόδων κατάταξης που θα περιγράψουμε στη συνέχεια. Αρχικά, εξετάζουμε τα μη αρνητικά μητρώα παρουσιάζοντας αναλυτικά τη θεωρία \en Perron-Frobenius \cite{langville2011google, meyer2000matrix} \gr και έπειτα κάνουμε μια συζήτηση περί πρωταρχικών μητρώων. Στη συνέχεια, αναφερόμαστε σε αλυσίδες \en Markov \gr \cite{mitzenmacher2005probability} παραθέτοντας κάποιους απαραίτητους ορισμούς. Τέλος, εξηγούμε σύντομα την έννοια της  \textit{Σχεδόν Πλήρους Αναλυσιμότητας} \en \textit{(Near Complete Decomposability} \gr \textit{ή} \en \textit{NCD)} \cite{courtoisdecomposability}. \gr

Στο \textbf{κεφάλαιο 3} περιγράφουμε διαισθητικά και μαθηματικά τους αλγορίθμους κατάταξης \en PageRank \gr \cite{page1999pagerank,brin1998anatomy} και \en NCDawareRank \gr \cite{nikolakopoulos2013ncdawarerank, RandomSurfingWithoutTeleportation}, οι οποίοι αποτελούν την βάση της μεθόδου που θα προτείνουμε, και ορίζουμε τα μητρώα που εμπλέκονται στον υπολογισμό των διανυσμάτων κατάταξης. 

Στο \textbf{κεφάλαιο 4} εκμεταλλευόμαστε την \en block \gr δομή των διμερών γραφημάτων μέσω του μητρώου τηλεμεταφοράς  και προτείνουμε έναν νέο αλγόριθμο, τον αλγόριθμο \en BipartiteRank\gr, προκειμένου να βελτιώσουμε τα αποτελέσματα που εξάγονται υπολογιστικά αλλά και ποιοτικά. 

Στο \textbf{κεφάλαιο 5} εκτελούμε μια σειρά υπολογιστικών πειραματικών μετρήσεων σε πραγματικά δεδομένα και επιχειρούμε άμεσες συγκρίσεις με τον αλγόριθμο \en PageRank\gr. 

Τέλος, στο \textbf{κεφάλαιο 6} εξάγουμε κάποια γενικότερα συμπεράσματα μαζί με προτάσεις για μελλοντική εργασία.

Σημειώνουμε πως τα σχήματα στα οποία στα οποία δεν αναφέρεται η πηγή έχουν κατασκευαστεί για τις ανάγκες αυτής της διπλωματικής με τη χρήση πραγματικών δεδομένων.

\chapter{Μαθηματικό Υπόβαθρο}

Για να γίνουν κατανοητές οι μαθηματικές λεπτομέρειες των αλγορίθμων με τους οποίους θα ασχοληθούμε στη συνέχεια, είναι απαραίτητο να συνοψίσουμε τα βασικά στοιχεία της θεωρίας που θα χρησιμοποιήσουμε. Το μεγαλύτερο τμήμα του κεφαλαίου αφιερώνεται στη θεωρία \en Perron-Frobenius \cite{langville2011google} \gr για μη αρνητικά μητρώα και στα κριτήρια που εξασφαλίζουν πως ένα μητρώο είναι πρωταρχικό. Έπειτα δίνονται κάποια βασικά στοιχεία και ορισμοί της θεωρίας των αλυσίδων \en Markov \cite{mitzenmacher2005probability, langville2011google} \gr και τέλος, γίνεται μια  σύντομη αναφορά στις \en Nearly Completely Decomposable \gr αλυσίδες \en Markov  \cite{courtoisdecomposability}. \gr

Στόχος μας σε αυτό το κεφάλαιο, δεν είναι να παρουσιάσουμε αναλυτικά όλα αυτά τα γνωστικά αντικείμενα, άλλα να θίξουμε εν συντομία μόνο τα σημαντικότερα μαθηματικά ζητήματα που θα φανούν χρήσιμα στον αναγνώστη.  

\section{Μη αρνητικά Μητρώα} \label{sec:nonnegativematrices}
Ένα μητρώο $\mathbf{Α}$ θεωρείται \textit{μη αρνητικό}, και γράφεται $\mathbf{Α} \ge 0$, αν τα στοιχεία του είναι μη αρνητικοί πραγματικοί αριθμοί. Αντίστοιχα, ένα μητρώο θεωρείται \textit{θετικό} και γράφεται $\mathbf{Α} >0$, αν τα στοιχεία του είναι θετικοί πραγματικοί αριθμοί. Σημαντικά παραδείγματα μη αρνητικών μητρώων είναι τα \textit{μητρώα γειτνίασης} γραφημάτων και τα \textit{στοχαστικά μητρώα}, τα οποία μας ενδιαφέρουν κατεξοχήν σε αυτή την διπλωματική λόγω του ότι χρησιμοποιούνται για να περιγράψουν αλυσίδες \en Markov\gr. 

\begin{definition} [Στοχαστικό Μητρώο]
Στοχαστικό μητρώο είναι ένα μη αρνητικό τετραγωνικό μητρώο $\mathbf{A}$ του οποίου το άθροισμα των στοιχείων κάθε γραμμής ισούται με 1. 
\end{definition}
 
Οι ιδιοτιμές ενός πραγματικού τετραγωνικού μητρώου $\mathbf{Α}$ είναι μιγαδικοί αριθμοί που συνθέτουν το φάσμα του μητρώου.  Το θεώρημα \en Perron-Frobenius\gr, το οποίο θα αναφέρουμε στη συνέχεια, περιγράφει τις ιδιότητες της επικρατούς ιδιοτιμής και του αντίστοιχου ιδιοδιανύσματος όταν το $\mathbf{Α}$ είναι ένα μη αρνητικό τετραγωνικό μητρώο πραγματικών αριθμών.

\subsection{Θετικά Μητρώα}
Σε αυτό το σημείο θα παρουσιάσουμε το θεώρημα του \en Perron \gr (1907), το οποίο αποτελεί τη βάση για την κατανόηση των ιδιοτήτων της επικρατούς ιδιοτιμής των θετικών μητρώων. Η απόδειξη του παρακάτω θεωρήματος περιλαμβάνεται στο εγχειρίδιο \cite{meyer2000matrix}.

	\begin{theorem}[Θεώρημα \en Perron \gr για Θετικά Μητρώα]
	Αν $\mathbf{Α}_{n \times n} > 0$ με $\lambda = \rho (\mathbf{Α})$, τότε ισχύουν τα ακόλουθα:
	\begin{enumerate}
	\item $\lambda \in \mathbb{R}$ και $\lambda>0$.
	\item $\lambda \in \sigma(\mathbf{Α})$  (η $\lambda$ ονομάζεται ρίζα \en Perron). \gr
	\item H ρίζα \en Perron \gr $\lambda$ είναι απλή.
	\item Υπάρχει θετικό ιδιοδιάνυσμα $\boldsymbol{x}>0$ τέτοιο ώστε 
	$\mathbf{Α}\boldsymbol{x}=\lambda \boldsymbol{x}$ (διάνυσμα \en Perron). \gr
	\item Το διάνυσμα \en Perron \gr είναι το μοναδικό διάνυσμα που ορίζεται από τη σχέση $\mathbf{Α}\boldsymbol{p}=\lambda \boldsymbol{p}$, όπου $\boldsymbol{p}>0$ και $||\boldsymbol{p}||_1=1$, και αν εξαιρέσουμε τα θετικά πολλαπλάσια του $\boldsymbol{p}$, δεν υπάρχουν άλλα μη αρνητικά ιδιοδιανύσματα του $\mathbf{Α}$, ανεξαρτήτως ιδιοτιμής. Όλα τα υπόλοιπα ιδιοδιανύσματα θα έχουν τουλάχιστον ένα αρνητικό ή μη πραγματικό στοιχείο.
	\item Η $\lambda$ είναι η μοναδική ιδιοτιμή επάνω στον φασματικό κύκλο του $\mathbf{Α}$.
	\item Tύπος \en Collatz-Wielandt \gr: $\lambda = \max_{x \in N } f(x)$, \begin{center}
	όπου $f(x) = \min_{1 \le i \le n}  \frac{[\mathbf{Α}x]_i} {x_i}, x_i \neq 0 $ και $N={x|x \ge 0, x \neq 0}$.
	\end{center}
	\end{enumerate}
	\label{the:perron}
	\end{theorem}	
	
Το θεώρημα \en Perron \gr  για θετικά μητρώα είναι εξαιρετικά χρήσιμο αποτέλεσμα. Θα ήταν φυσικό να αναρωτηθεί κανείς τι συμβαίνει όταν υπεισέρχονται στο μητρώο και κάποια μηδενικά στοιχεία. Σύμφωνα με το ακόλουθο θεώρημα (του οποίου η απόδειξη περιλαμβάνεται επίσης στο \cite{meyer2000matrix}), ένα τμήμα του θεωρήματος \en Perron \gr \ref{the:perron} για τα θετικά μητρώα μπορεί να επεκταθεί και στα μη αρνητικά μητρώα, αν θυσιάσουμε την ύπαρξη του θετικού ιδιοδιανύσματος και να αρκεστούμε σε ένα μη αρνητικό.
	\begin{theorem}
	Αν $\mathbf{Α}_{n \times n} \ge 0$ με φασματική ακτίνα $\lambda = \rho (\mathbf{Α})$, τότε ισχύουν τα ακόλουθα.
	\begin{itemize}
	\item $\lambda \in \sigma(\mathbf{Α})$ και $\lambda \ge 0$.
	\item Υπάρχει ιδιοδιάνυσμα $\boldsymbol{x} \ge 0$ τέτοιο ώστε $\mathbf{Α}\boldsymbol{x}=\lambda \boldsymbol{x}$.
	\item Ο τύπος \en Collatz-Wielandt \gr εξακολουθεί να ισχύει.
	\end{itemize}
	\end{theorem}
	
Το θεώρημα \en Perron \gr $\ref{the:perron}$ δεν είναι δυνατόν να επεκταθεί περισσότερο για μη αρνητικά μητρώα χωρίς κάποιες επιπλέον παραδοχές. Το 1912 ο \en F. G. Frobenius\gr, αντιλήφθηκε ότι το πρόβλημα αυτό δεν οφείλεται τόσο στην ύπαρξη μηδενικών στοιχείων, όσο στη θέση τους. Για παράδειγμα, η τρίτη και τέταρτη ιδιότητα του θεωρήματος \en Perron \gr δεν ισχύουν για το μητρώο $\bigl(\begin{smallmatrix} 1&0\\ 1&1 \end{smallmatrix} \bigr)$, ισχύουν όμως για το $\bigl(\begin{smallmatrix} 1&1\\ 1&0 \end{smallmatrix} \bigr)$. Σύμφωνα με τον \en Frobenius \gr η διαφορά μεταξύ των δύο μητρώων έγκειται στην μειωσιμότητα \en (reducibility) \gr ή μη μειωσιμότητά τους. \cite[σ.~167]{langville2011google}

\subsection{Μη μειωσιμότητα και Μη Περιοδικότητα}

Όπως ήδη γνωρίζουμε, ένα κατευθυνόμενο γράφημα ονομάζεται \textit{ισχυρά συνδεδεμένο}, αν για κάθε ζεύγος κορυφών $(i,j)$ υπάρχει μια ακολουθία ακμών που οδηγεί από την $i$ στην $j$. Σε κάθε κατευθυνόμενο γράφημα αντιστοιχεί ένα μητρώο, το οποίο ονομάζεται \textit{μητρώο γειτνίασης} του γραφήματος.

\begin{definition}[Αμείωτο Μητρώο]
Ένα οποιοδήποτε τετραγωνικό μητρώο $\mathbf{Α}$ είναι αμείωτο αν και μόνο αν το κατευθυνόμενο γράφημά του είναι ισχυρά συνδεδεμένο. Ισοδύναμα, το $\mathbf{Α}$ είναι αμείωτο αν για οποιοδήποτε μητρώο μετάθεσης $\mathbf{P}$ ισχύει ότι 
\begin{center}
 \[ \mathbf{P}^\top \mathbf{Α} \mathbf{P} \neq
\begin{bmatrix}
    \mathbf{X}  &  \mathbf{Y}   \\
    \mathbf{0}  &  \mathbf{Z}   \\
\end{bmatrix}
\]\end{center}
όπου τα μητρώα  $\mathbf{X}$ και $\mathbf{Z}$ είναι τετραγωνικά.
\end{definition}
	
Αν ένα \textit{μη αρνητικό μητρώο} $\mathbf{A}$ είναι \textit{αμείωτο}, για κάθε ζεύγος $i$ και $j$ υπάρχει ένας ακέραιος $t$ τέτοιος ώστε $\mathbf{A}^t_{i,j} > 0$. Για ένα  μητρώο γειτνίασης ενός κατευθυνόμενου γραφήματος, αυτή η ιδιότητα σημαίνει ότι το γράφημα είναι ισχυρά συνδεδεμένο. Αντίθετα, ένα μειώσιμο μητρώο γειτνίασης, αναπαριστά ένα γράφημα με περισσότερα από ένα ισχυρά συνδεδεμένα τμήματα.

\begin{definition}[Περίοδος Κορυφής]
Η περίοδος μιας κορυφής ενός γραφήματος ορίζεται ως ο μέγιστος κοινός διαιρέτης των μηκών όλων των κύκλων που περιλαμβάνουν την κορυφή.
\end{definition}

Όσον αφορά στην περιοδικότητα μιας κορυφής, αν υπάρχει μονοπάτι μήκους $t$ από μία κορυφή  $i$ στον εαυτόν της, τότε θα ισχύει $\mathbf{A}^t_{i,i}>0$.  Ο μέγιστος κοινός διαιρέτης του συνόλου $\{t: \mathbf{A}^t_{i,i}>0\}$ καλείται \textit{περίοδος} της κορυφής $i$. Αν το $\mathbf{A}$ είναι αμείωτο, τότε η περίοδος είναι η ίδια για όλες τις κορυφές και η κοινή αυτή περίοδος είναι η περίοδος του γραφήματος. Όπως είναι προφανές, ένα γράφημα με περίοδο μονάδα ονομάζεται \textit{μη περιοδικό} και το ίδιο ισχύει και για το μητρώο γειτνίασης του. \cite[σ.~128]{baldi2003modeling}

\subsection{Το Θεώρημα $\mathbf{Perron-Frobenius}$}

Όπως αναφέραμε και προηγουμένως, ο \en Frobenius \gr αντιλήφθηκε ότι παρόλο που οι ιδιότητες 1,2,4 και 6 του θεωρήματος του \en Perron \gr $\ref{the:perron}$ για θετικά μητρώα πιθανόν να πάψουν να ισχύουν για μη αρνητικά μητρώα, το πρόβλημα δεν είναι τόσο η παρουσία των μηδενικών στοιχείων όσο η θέση τους. Με άλλα λόγια, οι ιδιότητες 1,3 και 4 στην πραγματικότητα εξακολουθούν να ισχύουν αν τα μηδενικά βρίσκονται στις κατάλληλες θέσεις, δηλαδή σε θέσεις που να εξασφαλίζουν ότι το μητρώο είναι αμείωτο. Ωστόσο, δυστυχώς, η μη μειωσιμότητα από μόνη της δεν αρκεί για να διατηρηθεί η ιδιότητα 6.
\par Παρακάτω διατυπώνεται το θεώρημα \en Perron-Frobenius\gr. Η απόδειξη του θεωρήματος περιλαμβάνεται στο εγχειρίδιο του \en Carl Meyer \gr \cite{meyer2000matrix}.
\begin{theorem}[Θεώρημα \en Perron-Frobenius] \gr
Αν ένα μητρώο $\mathbf{A}_{n \times n} \ge 0$ είναι αμείωτο, τότε ισχύουν όλα τα ακόλουθα:
\begin{enumerate}
\item $\lambda=\rho(\mathbf{A}) > 0$, $\lambda \in \mathbb{R}$.
\item $\lambda \in \sigma(\mathbf{A})$ (η $\lambda$ ονομάζεται ρίζα \en Perron). \gr
\item Η ρίζα $\lambda$ είναι απλή.
\item Υπάρχει ιδιοδιάνυσμα $\boldsymbol{x}>0$ τέτοιο ώστε $\mathbf{A}\boldsymbol{x}=\lambda \boldsymbol{x}$.
\item Το διάνυσμα \en Perron \gr είναι το μοναδικό  διάνυσμα που ορίζεται από τη σχέση $\mathbf{A}\boldsymbol{p}=\lambda \boldsymbol{p}$, $\boldsymbol{p}>0$ και $||\boldsymbol{p}||_1=1$, και, αν εξαιρέσουμε τα θετικά πολλαπλάσια του $\boldsymbol{p}$, δεν υπάρχουν άλλα μη αρνητικά ιδιοδιανύσματα του $\mathbf{A}$, ανεξαρτήτως ιδιοτιμής.
\item Η $\lambda$ ΔΕΝ είναι απαραιτήτως η μοναδική ιδιοτιμή επάνω στο φασματικό κύκλο του $\mathbf{A}$.
\item Τύπος \en Collatz-Wielandt\gr: $\lambda = \max_{x \in N } f(x)$, \begin{center}
	όπου $f(x) = \min_{1 \le i \le n}  \frac{[\mathbf{A}x]_i} {x_i}, x_i \neq 0 $ και $N={x|x \ge 0, x \neq 0}$.
\end{center}
\end{enumerate}
\label{the:perron-frobenius}
\end{theorem}

Η $\lambda$ καλείται \textit{επικρατής ιδιοτιμή} του $\mathbf{A}$. Θα συμβολίσουμε με $(\lambda_1, ..., \lambda_n)$ τις ιδιοτιμές του $\mathbf{A}$, και στη συνέχεια θα θεωρούμε ότι η επικρατής ιδιοτιμή είναι πάντα η $\lambda_1$. Σημειώνουμε πως παρόλο που σύμφωνα με την πρόταση 3 του θεωρήματος \en Perron-Frobenis \gr $\ref{the:perron-frobenius}$ η ρίζα $\lambda_1$ είναι απλή, σύμφωνα με την πρόταση 6 μπορούν να υπάρχουν άλλες ιδιοτιμές $\lambda_j \neq \lambda_1$ τέτοιες ώστε $|\lambda_j| = |\lambda_1|$. Μπορεί να αποδειχθεί ότι αν υπάρχουν $k$ ιδιοτιμές, ίσες κατά μέτρο με την κύρια ιδιοτιμή, τότε είναι ομοιόμορφα τοποθετημένες σε έναν κύκλο ακτίνας $\lambda_1$. Επιπλέον, αν το μητρώο $A$ είναι το μητρώο γειτνίασης ενός γραφήματος, το $k$ είναι ο μέγιστος κοινός διαιρέτης των μηκών όλων των κύκλων του γραφήματος, δηλαδή το $k$ θα είναι η περίοδος του γραφήματος. 

Όπως είναι φανερό, προκειμένου να έχουμε μια κύρια ιδιοτιμή αυστηρά μεγαλύτερη και συνεπώς να είναι η μοναδική ιδιοτιμή πάνω στον φασματικό κύκλο του $\mathbf{A}$, θα πρέπει το αντίστοιχο γράφημα να μην έχει περίοδο 1 ή αλλιώς να είναι \textit{μη περιοδικό}.

Το σύνολο των μη αρνητικών αμείωτων μητρώων χωρίζεται σε δύο σημαντικές κλάσεις, ανάλογα με τον αν έχουν μόνο μια ιδιοτιμή επάνω στον φασματικό του κύκλο ή περισσότερες.
\begin{theorem}[Πρωταρχικά Μητρώα]\
\begin{itemize}
\item Το μητρώο $\mathbf{A}$ ονομάζεται \textit{πρωταρχικό} αν είναι ένα μη αρνητικό, αμείωτο μητρώο το οποίο έχει μόνο μια ιδιοτιμή, την $\lambda= \rho(\mathbf{A})$, επάνω στον φασματικό του κύκλο.
\item Ένα μη αρνητικό, αμείωτο μητρώο που έχει $h>1$ ιδιοτιμές επάνω στον φασματικό του κύκλο ονομάζεται μη πρωταρχικό, ενώ ο δείκτης $h$ ονομάζεται \textit{δείκτης μη πρωταρχικότητας}.
\item Αν το $\mathbf{A}$ είναι μη πρωταρχικό, τότε οι $h$ ιδιοτιμές επάνω στον φασματικό κύκλο είναι οι 
\begin{center}
$\{\lambda,\lambda \omega, \lambda \omega^2,...,\lambda \omega^{h-1}\}$, όπου $\omega=2e^{2\pi i /h}$.
\end{center}
Με άλλα λόγια, οι ιδιοτιμές είναι οι $h$-οστές ρίζες του $\lambda=\rho(\mathbf{A})$, και είναι ομοιόμορφα κατανεμημένες επάνω στον κύκλο. Επιπλέον, κάθε ιδιοτιμή $\lambda \omega^k$ επάνω στον φασματικό κύκλο είναι απλή.
\end{itemize}
\label{the:primitivity}
\end{theorem}

Γιατί όμως είναι τόσο σημαντικό να υπάρχει μόνο μια ιδιοτιμή  επάνω στον φασματικό κύκλο$;$ Η σημασία της πρωταρχικότητας έγκειται στο ότι είναι ακριβώς η ιδιότητα που καθορίζει κατά πόσο οι διαδοχικές δυνάμεις ενός κανονικοποιημένου μη αρνητικού αμείωτου μητρώου έχουν κάποια σταθερή-οριακή τιμή, στοιχείο από το οποίο εξαρτάται και η ύπαρξη του διανύσματος \en PageRank\gr. \cite{langville2011google, baldi2003modeling}

\section{Αλυσίδες $\mathbf{Markov}$}
Η μαθηματική συνιστώσα του διανύσματος \en PageRank \gr είναι η σταθερή κατανομή μιας αλυσίδας \en Markov \gr διακριτού χρόνου και πεπερασμένων καταστάσεων. Στη συνέχεια αναφέρουμε κάποιες πολύ χρήσιμες έννοιες.

\begin{definition}[Στοχαστική Διαδικασία]
  Μία στοχαστική διαδικασία  $X=\{ X_t:t \in T \}$ είναι μια συλλογή τυχαίων μεταβλητών. Ο δείκτης $t$ συνήθως αντιπροσωπεύει τον χρόνο. Η διαδικασία $Χ$ μοντελοποιεί την τιμή μιας τυχαίας μεταβλητής $X$ η οποία αλλάζει με το πέρασμα του χρόνου.
\end{definition}

Συμβολίζουμε $X_t$ την κατάσταση της διαδικασίας την χρονική στιγμή $t$. Για παράδειγμα, έστω μια διαδικασία ενός τυχαίου περιπάτου στον Παγκόσμιο Ιστό. Ο χώρος καταστάσεων είναι ο χώρος όλων των σελίδων ενώ η τυχαία μεταβλητή $X_t$ είναι η σελίδα στην οποία βρίσκεται ο περιηγητής τη χρονική στιγμή $t$. Για να δηλώσουμε ότι ο χρόνος δεν θεωρείται συνεχής αλλά διακριτός, χρησιμοποιούμε τον όρο \textit{διαδικασία διακριτού χρόνου}, ενώ για να δηλώσουμε ότι ο χώρος καταστάσεων είναι πεπερασμένος, χρησιμοποιούμε τον όρο \textit{διαδικασία πεπερασμένων καταστάσεων}. Σε αυτή την εργασία θα ασχοληθούμε με στοχαστικές διαδικασίες διακριτού χρόνου και πεπερασμένων καταστάσεων.

\begin{definition} [Αλυσίδα \en Markov] \gr
 Αλυσίδα \en Markov \gr είναι μια στοχαστική διαδικασία  που ικανοποιεί την ιδιότητα \en Markov \gr
 \begin{center}
 $Pr(X_t = a_t | X_{t-1} = a_{t-1}, X_{t-2}=a_{t-2}, ... ,X_0 = a_0) = Pr(X_t = a_t | X_{t-1}= a_{t-1}) = P_{a_{t-1},a_t}$  για κάθε $t=0,1,2,...$.
 \end{center}
\end{definition}

Η ιδιότητα \en  Markov \gr εκφράζει ότι η κατάσταση $X_t$ εξαρτάται από την προηγούμενη κατάσταση $X_{t-1}$ αλλά είναι ανεξάρτητη από το συγκεκριμένο ιστορικό καταστάσεων από το οποίο πέρασε η διαδικασία ώστε να φτάσει στην κατάσταση $X_{t-1}$, δηλαδή η στοχαστική διαδικασία δεν έχει μνήμη. Αν επανέλθουμε στο προηγούμενο παράδειγμα, η διαδικασία του τυχαίου περιπάτου είναι μια αλυσίδα \en Markov \gr μόνο εφόσον η σελίδα που επισκέπτεται κάθε φορά ο περιηγητής, δεν εξαρτάται από τις σελίδες που είχε επισκεφτεί προηγουμένως, αλλά αποκλειστικά από την τρέχουσα σελίδα. Με άλλα λόγια, αν ο περιηγητής επιλέγει τυχαία ένα \en link \gr της τρέχουσας σελίδας για να μεταβεί σε κάποια άλλη, τότε η διαδικασία είναι μια αλυσίδα \en Markov\gr.

Η \textit{πιθανότητα μετάβασης} η αλυσίδα να βρεθεί από την κατάσταση $i$ στην $j$ σε ένα βήμα τη χρονική στιγμή $t$ είναι
\begin{center} 
$\mathbf{P}_{i,j} = Pr(X_t = j | X_{t-1} = i)$
\end{center}
Επιπλέον, σύμφωνα με την \textit{ιδιότητα \en Markov\gr}, μια αλυσίδα \en Markov \gr ορίζεται από ένα \textit{μητρώο πιθανοτήτων μετάβασης} ενός βήματος:
\begin{center} 
\[ \mathbf{P}=
\begin{bmatrix}
    \mathbf{P}_{0,0}  &  \mathbf{P}_{0,1}  &  \dots  &  \mathbf{P}_{0,j}  &  \dots \\
    \mathbf{P}_{1,0}  &  \mathbf{P}_{1,1} &  \dots  &  \mathbf{P}_{1,j}  &  \dots \\
    \vdots   &  \vdots   &  \ddots &  \vdots   &  \ddots \\
    \mathbf{P}_{i,0}  &  \mathbf{P}_{i,1}  &  \dots  &  \mathbf{P}_{i,j} &  \dots \\
    \vdots   &  \vdots   &  \ddots &  \vdots   &  \ddots \\
\end{bmatrix}
\]
\end{center}
όπου για κάθε $i,j$ ισχύει
\begin{center} \begin{equation} 
\label{eq:transone}
\mathbf{P}_{i,j} \in [0,1]
\end{equation} \end{center}
και για κάθε $i$ ισχύει
\begin{center} \begin{equation}
\label{eq:transtwo}
\sum_{j \ge 0} \mathbf{P}_{i,j} =1 .
\end{equation} \end{center} 
Ένα μητρώο που ικανοποιεί τις εξισώσεις \ref{eq:transone} και \ref{eq:transtwo} είναι μη αρνητικό και το άθροισμα των στοιχείων κάθε γραμμής του είναι 1. Πρόκειται δηλαδή για ένα \textit{στοχαστικό μητρώο \en(stochastic matrix)\gr}. 

\begin{definition}[Ομογενής Αλυσίδα \en Markov] \gr
Ομογενής αλυσίδα \en Markov \gr είναι μια αλυσίδα \en Markov \gr στην οποία οι πιθανότητες μετάβασης δεν μεταβάλλονται  με τον χρόνο, όποτε οι πιθανότητες του $t$ βήματος μετάβασης μπορούν να υπολογιστούν σαν την $t$-οστη δύναμη του μητρώου πιθανοτήτων μετάβασης, $\mathbf{P}^t$.
\end{definition}

Στην περίπτωση αυτή, το μητρώο πιθανοτήτων μετάβασης είναι ένα στοχαστικό μητρώο $\mathbf{P}$. Στη συνέχεια της ανάλυσης αυτής, με τον όρο αλυσίδα \en Markov \gr θα αναφερόμαστε μόνο σε ομογενείς αλυσίδες \en Markov\gr. Κατά αυτόν τον τρόπο, μια αλυσίδα \en Markov \gr ορίζει μοναδικά ένα στοχαστικό μητρώο και το αντίστροφο.

\begin{definition}[Αμείωτη Αλυσίδα \en Markov]  \gr
Αμείωτη αλυσίδα \en Markov \gr είναι μια αλυσίδα \en Markov \gr για την οποία το μητρώο πιθανοτήτων μετάβασης $\mathbf{P}$ είναι αμείωτο. 
\end{definition}

\begin{definition} [Μη περιοδική Αλυσίδα \en Markov]  \gr
Μη περιοδική αλυσίδα \en Markov \gr είναι μια αμείωτη αλυσίδα της οποίας το μητρώο πιθανοτήτων μετάβασης είναι πρωταρχικό.
\end{definition}

Η περιοδικότητα οφείλεται στο γεγονός ότι κάθε κατάσταση επαναλαμβάνεται κατά περιοδικά χρονικά διαστήματα. Η περίοδος είναι ο δείκτης μη πρωταρχικότητας όπως αναφέρεται και στο θεώρημα \ref{the:primitivity}. Περιοδική είναι μια αμείωτη αλυσίδα \en Markov \gr της οποίας το μητρώο πιθανοτήτων μετάβασης $\mathbf{P}$ είναι μη πρωταρχικό.

Η κατανομή πιθανοτήτων των καταστάσεων μιας αλυσίδας \en Markov \gr μπορεί να αναπαρασταθεί με ένα \textit{διάνυσμα πιθανοτήτων κατάστασης} $\boldsymbol{p}^\top = (p_1,p_2,...,p_n)$, όπου όλα τα στοιχεία του ανήκουν στο διάστημα $[0,1]$, και έχουν άθροισμα 1. Κάθε στοιχείο του διανύσματος πιθανοτήτων κατάστασης αντιστοιχεί σε μία κατάσταση μιας αλυσίδας \en Markov\gr. 

\begin{definition}[Διάνυσμα Σταθερών Πιθανοτήτων Μετάβασης]
 Διάνυσμα σταθερών πιθανοτήτων κατάστασης μιας αλυσίδας \en Markov \gr με μητρώο πιθανοτήτων μετάβασης $\mathbf{P}$ είναι ένα διάνυσμα $\boldsymbol{\pi}^\top$ για το οποίο ισχύει η σχέση $\boldsymbol{\pi}^\top \mathbf{P} = \boldsymbol{\pi}^\top$.
\end{definition}

Ας επιστρέψουμε για ακόμη μια φορά στο παράδειγμα του τυχαίου περιπάτου. Στο βήμα $k=0$ ο περιηγητής μπορεί να ξεκινήσει από μια κατάσταση της οποίας το αντίστοιχο στοιχείο του $\boldsymbol{p}$ θα μπορούσε να είναι 1 για παράδειγμα, ενώ όλα τα υπόλοιπα είναι μηδέν. Η κατανομή πιθανοτήτων κατάστασης στο βήμα $k=1$ δίνεται από το διάνυσμα πιθανοτήτων $\boldsymbol{p}^\top \mathbf{P}$ και στο $k=2$ από το $(\boldsymbol{p}^\top \mathbf{P})\mathbf{P}$ και ούτω καθεξής. Με αυτόν τον τρόπο μπορούμε οποιαδήποτε στιγμή να υπολογίσουμε την κατανομή πιθανοτήτων κατάστασης σε οποιοδήποτε βήμα έχοντας ως δεδομένο μόνο την αρχική κατανομή και το μητρώο πιθανοτήτων μετάβασης $\mathbf{P}$. 

Αν επιτρέψουμε σε μία αλυσίδα \en Markov \gr να εκτελεστεί για πολλά χρονικά βήματα, κάθε κατάσταση θα προσπελαστεί με συχνότητα που εξαρτάται από τη δομή της αλυσίδας. Κατ' αναλογία, ο περιηγητής επισκέπτεται ορισμένες σελίδες (για παράδειγμα δημοφιλείς σελίδες ειδήσεων) συχνότερα από άλλες. Μας ενδιαφέρει λοιπόν, οι κατανομές πιθανοτήτων κατάστασης να μην αλλάζουν μετά από μία μετάβαση. Θα προχωρήσουμε τώρα στην επακριβή διατύπωση αυτής της λογικής, καθορίζοντας τις συνθήκες υπό τις οποίες η κατανομή πιθανοτήτων κατάστασης μιας αλυσίδας \en Markov \gr συγκλίνει σε μία συγκεκριμένη τιμή σταθερής κατάστασης. Όταν και αν η αλυσίδα φτάσει σε μία σταθερή κατανομή, τότε διατηρεί την κατανομή αυτή σε όλες τις μελλοντικές χρονικές στιγμές. \cite[σ.~167]{langville2011google}

\subsection{Εργοδικότητα} \label{ergodicity}

 Μια πεπερασμένων καταστάσεων αλυσίδα \en Markov \gr λέμε ότι είναι \textit{εργοδική} αν είναι \textit{αμείωτη} και \textit{μη περιοδική}.
\begin{theorem} \label{the:ergodicity}
Μια πεπερασμένων καταστάσεων, αμείωτη, και εργοδική αλυσίδα \en Markov \gr έχει τις παρακάτω ιδιότητες:
\begin{itemize}
\item η αλυσίδα έχει μια μοναδική σταθερή κατανομή πιθανοτήτων $\boldsymbol{\pi}^\top$.
\item για κάθε $j$ και $i$, το όριο $\lim_{t \rightarrow \infty} \mathbf{P}_{j,i}^t$ υπάρχει και είναι ανεξάρτητο του $j$.
%\item $\boldsymbol{\pi}_i= lim_{t \rightarrow \infty} \mathbf{P}_{ji}^t = \frac{1}{h_{i,i}}$
\end{itemize} \end{theorem}
\par Η εργοδικότητα είναι μια πολύ σημαντική ιδιότητα για το πρόβλημα κατάταξης που μας απασχολεί, αφού αν η αλυσίδα \en Markov \gr είναι εργοδική τότε ανεξάρτητα από τις αρχικές πιθανότητες, αυτή συγκλίνει σε ένα μοναδικό αναλλοίωτο μέτρο πιθανότητας. \cite[σ.~168]{mitzenmacher2005probability}

\section{Σχεδόν Πλήρης Αναλυσιμότητα} \label{subs:nearcompletedecomposability}
Τα συστήματα που μπορούν να χωριστούν σε \en blocks \gr με τις αλληλεπιδράσεις ανάμεσα στα \en blocks \gr να είναι μη μηδενικές, αλλά μικρές σε σύγκριση με τις αλληλεπιδράσεις εντός των \en blocks\gr, χαρακτηρίζονται ως \en Nearly Completely Decomposable (NCD)\gr. Οι \en Simon \gr και \en Ando \gr \cite{simon1961aggregation} ασχολήθηκαν πρώτοι με την ανάλυση \en NCD \gr συστημάτων και ακολούθησε ο \en Courtois \gr \cite{courtoisdecomposability}, ο οποίος θεμελίωσε και μαθηματικά την ιδέα της \en Decomposability\gr.

Στη θεωρία πιθανοτήτων, μια \en NCD \gr Αλυσίδα \en Markov \gr είναι μια αλυσίδα \en Markov \gr της οποίας ο χώρος καταστάσεων μπορεί να διαιρεθεί σε $K$ \en blocks \gr με τέτοιον τρόπο, ώστε οι μεταβάσεις μεταξύ καταστάσεων του ίδιου \en block \gr να είναι πιο πιθανές από τις μεταβάσεις μεταξύ καταστάσεων που ανήκουν σε διαφορετικά \en blocks\gr. Το στοχαστικό μητρώο πιθανοτήτων μετάβασης μιας \en NCD \gr αλυσίδας  \en Markov \gr μπορεί να εκφραστεί ως εξής
\begin{center}
\[ \mathbf{P}=
\begin{bmatrix}
    \mathbf{P}_{1,1}  &  \mathbf{P}_{1,2}  &  \dots  &  \mathbf{P}_{1,K}\\
    \mathbf{P}_{2,1}  &  \mathbf{P}_{2,2}  &  \dots  &  \mathbf{P}_{2,K}\\
    \vdots   &  \vdots   &  \ddots &  \vdots \\
    \mathbf{P}_{K,1}  &  \mathbf{P}_{K,2}  &  \dots  &  \mathbf{P}_{K,K}\\
\end{bmatrix}
\]
$||\mathbf{P}_{i,i}|| = O(1), i=1,2,...,K$ \\ $||\mathbf{P}_{i,j}|| = O(\epsilon), i \ne j$
\end{center}
όπου $\epsilon$ είναι ένας ικανοποιητικά μικρός θετικός αριθμός.

Θεωρούμε ένα $n \times n$ αμείωτο στοχαστικό μητρώο $\mathbf{P}$, που αναπαριστά το μητρώο πιθανοτήτων μετάβασης μιας εργοδικής αλυσίδας \en Markov\gr. Θεωρούμε επίσης, τη στοχαστική διαδικασία $y_t$ με $t \in \mathbb{N}$. Τα συστήματα που μας ενδιαφέρουν έχουν την μορφή
\begin{center}
$y^\top_{t+1}= y^\top_{t} \mathbf{P}$
\end{center}
Το $\mathbf{P}$ μπορεί να γραφεί ως εξής
\begin{center} \begin{equation} \label{eq:ncdone}
\mathbf{P} = \mathbf{P}^\ast + \zeta \mathbf{C}
\end{equation}\end{center}
όπου $\mathbf{P}^\ast$ είναι ένα \en block\gr-διαγώνιο μητρώο τάξης $n$ που δίνεται από
\begin{center}
$\mathbf{P}^\ast=\mathbf{\operatorname{Diag}}(\mathbf{P}^\ast_{1},\mathbf{P}^\ast_{2}...\mathbf{P}^\ast_{K})$
\end{center}
όπου τα μητρώα $\mathbf{P}^\ast_{i}, i=1,2,...,K$, είναι αμείωτα στοχαστικά μητρώα τάξης $n(i)$. Συνεπώς, τα αθροίσματα των γραμμών του μητρώου $\mathbf{C}$ είναι όλα μηδέν. Το μητρώο $\mathbf{C}$ και ο μη αρνητικός αριθμός $\zeta$ επιλέγονται με τέτοιον τρόπο, ώστε για κάθε γραμμή $m_i, i=1,2,...,K$ να ισχύει:
\begin{center}\begin{equation} \label{eq:ncdtwo}
\zeta \sum_{j \neq i} \sum^{n_j}_{i=1} \mathbf{C}_{m_i, i_j} = \sum_{j \neq i} \sum^{n_j}_{i=1}\mathbf{P}_{m_i, i_j}
\end{equation}\end{center} και
\begin{center}\begin{equation} \label{eq:ncdthree}
\zeta = \max_{m_i}(\sum_{j \neq i} \sum^{n_j}_{i=1}\mathbf{P}_{m_i, i_j})
\end{equation}\end{center} 
όπου $\mathbf{P}_{m_i i_j}$, συμβολίζει το στοιχείο στην τομή της $m$-οστής γραμμής και $i$-οστής στήλης του υπομητρώου $\mathbf{P}_{i j}$ του $\mathbf{P}$. Η παράμετρος $\zeta$ αναφέρεται, ως ο μέγιστος βαθμός σύζευξης μεταξύ των υποσυστημάτων $\mathbf{P}^\ast_{i i}$.

Ένα παράδειγμα των μητρώων $\mathbf{P}$, $\mathbf{P^\ast}$ και $\mathbf{C}$ μπορεί να είναι το εξής
\begin{center}
\[ \mathbf{P}= 
\begin{bmatrix}
    0.5  &  0.45  & 0.05 \\
    0.6  &  0.375  &  0.025\\
    0.025 & 0.025 & 0.95\\
\end{bmatrix},
\]\end{center} 
με τη σχέση $\mathbf{P} = \mathbf{P^\ast} + \zeta \mathbf{C}$ να δίνεται από το:
\begin{center}
\[ \mathbf{P}= 
\begin{bmatrix}
    0.5  &  0.5  & 0 \\
    0.625  &  0.375  &  0\\
    0 & 0 & 1\\
\end{bmatrix} + 5 \times 10^{-2}
\begin{bmatrix}
    0  &  -1  & 1 \\
    -0.5  &  0  &  0.5\\
    0.5 & 0.5 & -1\\
\end{bmatrix}
\]\end{center}
με  
\begin{center}
\[ \mathbf{P}^\ast_{1}= 
\begin{bmatrix}
    0.5  &  0.5  \\
    0.625  &  0.375\\
\end{bmatrix}, 
\mathbf{P}^\ast_{2}= 
\begin{bmatrix}
    1  \\
\end{bmatrix}\]\end{center}

Συμβολίζουμε με $\lambda^\ast(j_{i}), i = 1,...,n(i)$ τις ιδιοτιμές του $\mathbf{P}^\ast_{i}$ και υποθέτουμε πως μπορούμε να τις ταξινομήσουμε με τέτοιο τρόπο ώστε:
\begin{center}
$\lambda^\ast(1_{i}) = 1 >|\lambda^\ast(2_{i})| \ge |\lambda^\ast(2_{i})| ... \ge |\lambda^\ast(n(j)_{i})|$
\end{center} Οι ιδιοτιμές που δεν είναι ίσες με 1 θα πρέπει να είναι διακριτες. 

Στα \en NCD \gr συστήματα η δυναμική συμπεριφορά του $\mathbf{P}$ μπορεί να προσεγγισθεί από την μελέτη του μητρώου $\mathbf{P}^\ast$. Για να γίνει αυτό, θα πρέπει να μελετηθούν στοχαστικά συστήματα της μορφής $y^{\ast\top}_{t+1}= y^{\ast\top}_{t} \mathbf{P}^\ast$ και να εξεταστούν οι συνθήκες υπό τις οποίες το μονοπάτι που ακολουθεί η διαδικασία $y_0^\ast,y_1^\ast,y_1^\ast...$ συγκλίνει στο ακριβές $y_0,y_1,y_2,...$. Θα πρέπει δηλαδή να εξεταστεί η συμπεριφορά του συστήματος στον χρόνο. Για περισσότερες λεπτομέρειες σχετικά με την δυναμική συμπεριφορά των \en NCD \gr συστημάτων και την εξέλιξη τους στο χρόνο θα μπορούσε κάποιος να ανατρέξει στο \cite{courtoisdecomposability}.

\chapter{Αλγόριθμοι Κατάταξης} 

O αλγόριθμος \en NCDawareRank \cite{nikolakopoulos2013ncdawarerank,nikolakopoulos2015random}\gr, είναι ένας αλγόριθμος κατάταξης που εκμεταλλεύεται την ιεραρχική διάρθρωση του Παγκόσμιου Ιστού και πατά διαισθητικά στην ιδιότητα \en NCD\gr, προκειμένου να γενικεύσει και να βελτιώσει τόσο ποιοτικά όσο και υπολογιστικά τον αλγόριθμο \en PageRank \cite{brin1998anatomy, page1999pagerank}\gr.

Σε αυτό το κεφάλαιο, θα παρουσιάσουμε αρχικά τον μαθηματικό συλλογισμό του \en PageRank\gr, ο οποίος θα αποτελέσει στη συνέχεια τη βάση της μεθόδου μας. Έπειτα θα περιγράψουμε τον \en NCDawareRank\gr, από τον οποίο πηγάζει ουσιαστικά η κεντρική της ιδέα. Οι δύο αυτοί αλγόριθμοι είναι σχεδιασμένοι ώστε να εξάγουν ένα διάνυσμα κατάταξης μεταξύ των σελίδων του Παγκόσμιου Ιστού, και μπορούν να εφαρμοστούν και σε πολλά άλλα γραφήματα πραγματικού κόσμου.

\section{Ο αλγόριθμος $\mathbf{PageRank}$} \label{subs:pagerank}

\subsection{Το μοντέλο του Τυχαίου Περιηγητή} \label{RandomSurferModel}
Στη συνέχεια θα περιγράψουμε τον αλγόριθμο \en PageRank \gr ακολουθώντας τη συλλογιστική πορεία του βιβλίου \cite{langville2011google}. 

\begin{figure}
\centering
 \includegraphics[width=0.3\textwidth]{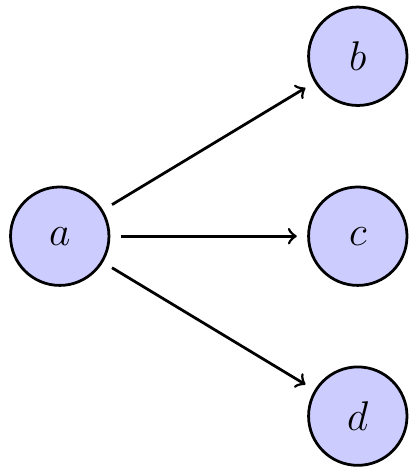}
\caption{Στιγμιότυπο τυχαίου περιπάτου}
\label{fig:randomwalkinstant}
\end{figure}

O αλγόριθμος \en PageRank\gr, των \en L. Page \gr και \en S. Brin\gr, συμπεριφέρεται σαν ένας τυχαίος περιηγητής, που μεταβαίνει από σελίδα σε σελίδα ακολουθώντας τυχαία διαδοχικά \en links\gr, χωρίς να τον απασχολεί το περιεχόμενο. Σε κάθε βήμα, μεταβαίνει από την τρέχουσα κορυφή $a$ σε μία τυχαία επιλεγμένη κορυφή προς την οποία δείχνει η $a$. Για παράδειγμα \cite[σ.~522]{manning2008introduction}, το σχήμα $\ref{fig:randomwalkinstant}$ παρουσιάζει τον τυχαίο περιηγητή στην κορυφή $a$ όπου μπορεί να ακολουθήσει μία από τις τρεις ακμές προς τις κορυφές $b$, $c$ και $d$. Στο επόμενο χρονικό βήμα θα προχωρήσει σε μία από τις τρεις κορυφές με πιθανότητες $1/3$ για την καθεμία.

Με βάση αυτή τη λογική, μπορούμε να κατασκευάσουμε ένα μητρώο πιθανοτήτων μετάβασης $\mathbf{H}$ το οποίο προκύπτει από την κανονικοποίηση κατά γραμμές του μητρώου γειτνίασης του γραφήματος Ιστού. Το μητρώο αυτό είναι υποστοχαστκό καθώς οι μη μηδενικές γραμμές του είναι στοχαστικές. Το $\mathbf{H}$ δίνεται από τον τύπο
\begin{center}
$ \mathbf{H}_{u,v} = \left\{
  \begin{array}{l l}
    \frac{1}{d_u} & \quad \text{αν υπάρχει ακμή απο το u στο v}\\
    0 & \quad \text{αλλιώς}
  \end{array} \right.$
\end{center}
όπου συμβολίζουμε με  $\mathcal{G}_u$ το σύνολο των κορυφών που μπορούμε να επισκεφθούμε με ένα βήμα από την $u$ και με $d_u$ τον βαθμό εξόδου της $u$.

Τι συμβαίνει όμως με τις μηδενικές γραμμές$;$ Τι συμβαίνει δηλαδή, όταν η κορυφή που βρίσκεται ο τυχαίος περιηγητής δεν έχει εξερχόμενες ακμές ή ακόμη όταν εγκλωβιστεί σε κάποιον βρόγχο$;$ Για να αντιμετωπίσουν ένα τέτοιο ενδεχόμενο οι \en L. Page \gr και \en S. Brin\gr, έδωσαν στον τυχαίο περιηγητή τη δυνατότητα τηλεμεταφοράς. Με την τηλεμεταφορά, μπορεί να μεταφερθεί από μια κορυφή σε οποιαδήποτε άλλη κορυφή του γραφήματος Ιστού. Αυτό θα μπορούσε να το καταφέρει στην πράξη πληκτρολογώντας απευθείας ένα \en URL \gr στη γραμμή διευθύνσεων του φυλλομετρητή του. Ο προορισμός μιας τηλεμεταφοράς επιλέγεται ομοιόμορφα και τυχαία από όλες τις σελίδες. Με άλλα λόγια, αν το πλήθος των κορυφών του γραφήματος είναι $n$, η τηλεμεταφορά μεταφέρει τον τυχαίο περιηγητή σε κάθε κορυφή (συμπεριλαμβανομένης και της τρέχουσας κορυφής) με πιθανότητα $1/n$.  

Στην ανάθεση της βαθμολογίας \en PageRank \gr σε κάθε κορυφή του γραφήματος Ιστού, η δυνατότητα της τηλεμεταφοράς χρησιμοποιείται με δύο τρόπους:
\begin{itemize}
  \item Όταν ο τυχαίος περιηγητής βρεθεί σε κορυφή χωρίς εξερχόμενες ακμές δεν έχει επιλογές μετάβασης. Άρα θα πρέπει να χρησιμοποιήσει τηλεμεταφορά. Για να γίνει αυτό εφικτό οι γραμμές του  $\mathbf{H}$ που είναι ίσες με $\boldsymbol{0}^\top$ θα πρέπει να αντικατασταθούν με $\frac{1}{n}\boldsymbol{e}^\top$, ώστε το  $\mathbf{H}$ να γίνει στοχαστικό μητρώο.
  \item Σε κάθε κορυφή με εξερχόμενες ακμές, ενεργοποιεί την τηλεμεταφορά με πιθανότητα $1-\epsilon$ ή συνεχίζει τον τυχαίο περίπατο με πιθανότητα $\epsilon$, όπου $\epsilon$ είναι μια σταθερή παράμετρος επιλεγμένη εκ των προτέρων και ισχύει $0 \le \epsilon \le 1$. Έτσι, θα εξασφαλιστεί και η πρωταρχικότητα του $\mathbf{H}$. 
\end{itemize}

Με βάση τις παραπάνω διορθώσεις στον τυχαίο περίπατο, θεωρούμε ένα μητρώο $\mathbf{S}$ που εκφράζει το $\mathbf{H}$ μετά την αντικατάσταση των $\boldsymbol{0}^\top$ γραμμών του, δηλαδή $\mathbf{S}=\mathbf{Η}+a(\frac{1}{n}\boldsymbol{e}^\top)$, όπου $a_i=1$ αν η κορυφή $i$ δεν έχει εξερχόμενες ακμές και ένα ομοιόμορφο μητρώο τηλεμεταφοράς $\mathbf{E}=\frac{1}{n}\boldsymbol{e} \boldsymbol{e}^\top$. Το τελικό μητρώο που προκύπτει από τον συνδυασμό τυχαίου περιπάτου και τηλεμεταφοράς, το λεγόμενο μητρώο της \en Google \gr \cite[σ.~37]{langville2011google}, είναι το εξής
\begin{center}
\begin{equation}
\mathbf{G} = \epsilon \mathbf{S} + (1-\epsilon) \mathbf{E}.
\end{equation}
\end{center}

\subsection{Ο υπολογισμός του διανύσματος $\mathbf{PageRank}$}
Η προσαρμογή πρωταρχικότητας που περιγράψαμε στην προηγούμενη υποενότητα έχει τις εξής συνέπειες \cite[σ.~38]{langville2011google}: 
\begin{itemize}
\item Το $\mathbf{G}$ είναι στοχαστικό. Αποτελεί κυρτό μητρώο δύο στοχαστικών μητρώων $\mathbf{S}$ και $\mathbf{E}$.
\item Το $\mathbf{G}$ είναι αμείωτο. Η μη μειωσιμότητα εξασφαλίζεται άμεσα, αφού κάθε κορυφή συνδέεται άμεσα με όλες τις άλλες.
\item Το $\mathbf{G}$ είναι μη περιοδικό. Η μη περιοδικότητα οφείλεται στους ιδιοβρόγχους του γραφήματος.
\item Το $\mathbf{G}$ είναι πρωταρχικό, διότι υπάρχει ακέραιος $k$ τέτοιος ώστε $\mathbf{G}^k >0 $ (και μάλιστα αυτό ισχύει ήδη για $k=1$). Αυτό συνεπάγεται ότι υπάρχει ένα μοναδικό θετικό διάνυσμα $\boldsymbol{\pi}^\top$ και ότι αν μια δυναμομέθοδος εφαρμοστεί στο $\mathbf{G}$, θα συγκλίνει σίγουρα σε αυτό το διάνυσμα.
\item Το $\mathbf{G}$ είναι απολύτως πυκνό, πράγμα ιδιαίτερα δυσάρεστο από υπολογιστικής πλευράς. Ευτυχώς, όμως, το $\mathbf{G}$ μπορεί να γραφτεί ως ενημέρωση πρώτης τάξης στο πολύ αραιό μητρώο υπερσυνδέσμων $\mathbf{H}$.
\begin{center}\begin{equation}
\begin{split}
\mathbf{G}  & = \epsilon \mathbf{S} + (1-\epsilon) \frac{1}{n}\boldsymbol{e} \boldsymbol{e}^\top \\
& = \epsilon (\mathbf{H} + \frac{1}{n}a\boldsymbol{e}^\top) + (1 - \epsilon) \frac{1}{n} \boldsymbol{e} \boldsymbol{e}^\top \\
& = \epsilon \mathbf{H} +(\epsilon a + (1-\epsilon)\boldsymbol{e})\frac{1}{n} \boldsymbol{e}^\top.
\end{split}
\end{equation}\end{center} 
\end{itemize}
\par Εν ολίγοις, η προσαρμοσμένη μέθοδος \en PageRank \gr της \en Google \gr εκφράζεται από την επαναληπτική εξίσωση
\begin{center}\begin{equation} \label{powermethodone}
\boldsymbol{\pi^{(k+1)\top}}=\boldsymbol{\pi^{(k)\top}} \mathbf{G}
\end{equation}\end{center} 
δηλαδή είναι μια δυναμομέθοδος εφαρμοσμένη στο μητρώο της \en Google\gr. 

Η δυναμομέθοδος είναι μια από τις παλαιότερες και απλούστερες επαναληπτικές μεθόδους εύρεσης των επικρατών ιδιοτιμών και ιδιοδιανυσμάτων ενός μητρώου και επομένως μπορεί να χρησιμοποιηθεί για την εύρεση του σταθερού διανύσματος μιας αλυσίδας \en Markov \gr (το σταθερό διάνυσμα δεν είναι τίποτα άλλο παρά το επικρατές αριστερό ιδιοδιάνυσμα του μητρώου πιθανοτήτων μετάβασης). Ωστόσο, είναι εξαιρετικά αργή και η πιο αργή σε σχέση με άλλες επαναληπτικές μεθόδους (\en Gauss-Seidel, Jacobi \cite{stewart2009probability}, \gr και λοιπά). 

Οι λόγοι επιλέχθηκε η μέθοδος αυτή, είναι γιατί είναι πολύ απλή, και επιπλέον αν εφαρμοστεί στο $\mathbf{G}$, όπως στην εξίσωση \ref{powermethodone}, μπορεί τελικά να εκφραστεί μέσω του πολύ αραιού μητρώου $\mathbf{H}$
\begin{center}\begin{equation} \label{powermethodtwo}
\begin{split}
\boldsymbol{\pi^{(k+1)\top}}  & = \boldsymbol{\pi^{(k)\top}} \mathbf{G}\\
& = \epsilon \boldsymbol{\pi^{(k)\top}} \mathbf{S} + (1-\epsilon) \frac{1}{n} \boldsymbol{\pi^{(k)\top}} \boldsymbol{e} \boldsymbol{e}^\top \\
& = \epsilon \boldsymbol{\pi^{(k)\top}} \mathbf{H} +(\epsilon \boldsymbol{\pi^{(k)\top}} a + (1-\epsilon))\frac{1}{n} \boldsymbol{e}^\top.
\end{split}
\end{equation}\end{center} 
Οι πολλαπλασιασμοί \en MV \gr ($\boldsymbol{\pi^{(k)\top}} \mathbf{H}$) αφορούν το ιδιαίτερα αραιό μητρώο $\mathbf{H}$. Τα $\mathbf{S}$ και $\mathbf{G}$ δεν χρειάζεται να υπολογιστούν ούτε να αποθηκευτούν. Το μόνο που χρειάζεται είναι τα διανύσματα τάξης ένα $a$ και $\boldsymbol{e}$ από τα οποία σχηματίζονται τα μητρώα αυτά. Υπενθυμίζουμε ότι κάθε πολλαπλασιασμός \en MV \gr έχει πολυπλοκότητα $O(n)$ \cite[σ.~40]{langville2011google}.

Όσον αφορά στη σύγκλιση, η δυναμομέθοδος όταν εφαρμόζεται στο μητρώο της \en Google \gr $\mathbf{G}$ απαιτεί μόνο 50 περίπου επαναλήψεις για να συγκλίνει. Υπάρχει κάτι στη δομή του $\mathbf{G}$ που δικαιολογεί αυτήν την ταχεία σύγκλιση και την εξήγηση τη δίνει η θεωρία \en Markov\gr. Όταν η δυναμομέθοδος εφαρμόζεται σε έναν μητρώο, ο ασυμπτωτικός ρυθμός σύγκλισης της εξαρτάται από τον λόγο των δύο μεγαλύτερων κατά μέτρο ιδιοτιμών, $\lambda_1$ και $\lambda_2$. Για την ακρίβεια ο ασυμπτωτικός ρυθμός σύγκλισης είναι ο ρυθμός με τον οποίο $|\lambda_2 / \lambda_1|^k \rightarrow 0$. Για στοχαστικά μητρώα, όπως το $\mathbf{G}$ έχουμε $\lambda_1=1$ και επομένως η σύγκλιση καθορίζεται από το $|\lambda_2|$. Δεδομένου ότι το $\mathbf{G}$ είναι πρωταρχικό, έχουμε $|\lambda_2|<1$. Ο αριθμητικός υπολογισμός της ιδιοτιμής $\lambda_2$ ενός μητρώου έχει μεγάλο υπολογιστικό κόστος και δεν αξίζει να υπολογιστεί μόνο και μόνο για να γίνει μια εκτίμηση του ασυμπτωτικού ρυθμού σύγκλισης.

\begin{theorem}[Υποεπικρατής ιδιοτιμή του μητρώου της \en  Google]  \gr
 Για το μητρώο της \en Google \gr $\mathbf{G}=\epsilon \mathbf{S} + (1-\epsilon) \frac{1}{n} \boldsymbol{e} \boldsymbol{e^\top}$,
 \begin{center}
 $|\lambda_2(\mathbf{G})| \le \epsilon$.
 \end{center}
 Σε περίπτωση που  $|\lambda_2(\mathbf{S})|=1 $ (το οποίο συμβαίνει συχνά λόγω της μειωσιμότητας του $\mathbf{S}$), έχουμε ότι $|\lambda_2(\mathbf{G})|= \epsilon$. Επομένως, ο ασυμπτωτικός ρυθμός σύγκλισης της δυναμομεθόδου \en PageRank \gr της εξίσωσης $\ref{powermethodtwo}$ είναι ο ρυθμός με τον οποίο $\epsilon^k \rightarrow 0$.
\label{the:subdominantpagerank}
\end{theorem}
Στα άρθρα τους \cite{brin1998anatomy,page1999pagerank}, οι \en S. Brin \gr και \en L. Page \gr χρησιμοποιούν την τιμή $\epsilon = 0.85$ και αιτιολογούν αναλυτικά την επιλογή τους αυτή.

\subsection{Αδυναμίες}

Παρόλο που ο \en PageRank \gr έχει αποδειχθεί πως είναι ένας από τους πιο αποδοτικούς αλγορίθμους κατάταξης στον Παγκόσμιο Ιστό παρουσιάζει αρκετές αδυναμίες. Μία από αυτές είναι η ευαισθησία του στο \en link spamming\gr, το οποίο περιλαμβάνει την σκόπιμη δημιουργία μεγάλου αριθμού σελίδων οι οποίες δείχνουν σε μία συγκεκριμένη σελίδα, με στόχο την τεχνητή ενίσχυση της βαθμολογίας της τελευταίας \cite{ntoulas2006detecting}. 

Άλλη μια αδυναμία, προέρχεται από το γεγονός ότι το γράφημα Ιστού είναι εξαιρετικά αραιό. Σύμφωνα με τις ερευνητικές εργασίες \cite{kleinberg1999web,barabasi1999emergence,broder2000graph} και άλλες μεταγενέστερες, ο αριθμός των υπερσυνδέσμων ανά σελίδα ακολουθεί έναν εκθετική νόμο, σύμφωνα με τον οποίο το συνολικό πλήθος των σελίδων με βαθμό εισόδου $i$ είναι ανάλογο προς το $1/i^\alpha$, με το $\alpha$ να αναφέρεται από τις περισσότερες μελέτες να είναι 2.1. Αυτό οδηγεί σε αρκετά αραιό μητρώο υπερσυνδέσμων. Επιπλέον, τέτοιου είδους κατανομές κάνουν τις πιθανότητες που παράγονται μέσω του \en PageRank \gr να μειώνονται πάλι σύμφωνα με τον εκθετικό νόμο, καθιστώντας αδύνατο σε ορισμένες σελίδες να αποκτήσουν μια λογική βαθμολογία, ειδικά όταν αυτές είναι νεοεισερχόμενες \cite{nikolakopoulos2013ncdawarerank}.

\section{Ο αλγόριθμος $\mathbf{NCDawareRank}$}\label{sec:ncdawarerank}

\subsection{Αξιοποιώντας την Σχεδόν Πλήρη Αναλυσιμότητα}

Ο στόχος των συγγραφέων των \cite{nikolakopoulos2013ncdawarerank, RandomSurfingWithoutTeleportation}, ήταν να παρουσιάσουν μια νέα προσέγγιση, που εκμεταλλεύεται το γεγονός ότι ο Παγκόσμιος Ιστός μπορεί να θεωρηθεί \en NCD \gr σύστημα, ώστε να  παράγουν καλύτερης ποιότητας κατάταξη σε σχέση με τον \en PageRank\gr. Η ιδέα πάνω στην οποία στηρίχτηκε ο \en NCDawareRank\gr, είναι ότι η ιεραρχική διάρθρωση του χώρου των σελίδων θα μπορούσε να απεικονιστεί και στο τελικό μητρώο. Για λόγους που περιγράφονται αναλυτικά στις ερευνητικές εργασίες \cite{nikolakopoulos2013ncdawarerank,kamvar2003exploiting,bharat2001links}, η βασική μονάδα διαίρεσης του χώρου των σελίδων είναι ο ιστότοπος, έτσι οι σελίδες που ανήκουν στον ίδιο ιστότοπο θεωρείται ότι αποτελούν ένα \en NCD block\gr.

Πιο συγκεκριμένα, γίνεται η υπόθεση πως ο τυχαίος  περιηγητής έχει μεγαλύτερη πιθανότητα να μεταβεί μέσω της γραμμής διευθύνσεων σε κάποια σελίδα που ανήκει στο ίδιο ή σε κάποιο από τα γειτονικά \en NCD blocks\gr. Με αυτόν τον τρόπο, η ύπαρξη εξερχόμενου υπερσυνδέσμου εκτός του ότι ενισχύει τη βαθμολογία της σελίδας προς την οποία δείχνει, ενισχύει και τη βαθμολογία των σελίδων που ανήκουν στο ίδιο και σε γειτονικά \en NCD blocks. \gr

Η ίδια λογική ακολουθείται στα \cite{nikolakopoulos2014ncdrec, nikolakopoulos2015hierarchical}, όπου οι ίδιοι συγγραφείς προτείνουν ένα \en top-N recommendation \gr πλαίσιο, το οποίο εκμεταλλεύεται την κρυμμένη ιεραρχική δομή του χώρου των αντικειμένων που εφαρμόζεται αντιμετωπίζοντας έτσι το πρόβλημα της αραιότητας. Μάλιστα, τονίζουν πως η πλειοψηφία των αραιών γραφημάτων του πραγματικού κόσμου κρύβει μια ιεραρχική δομή από \en NCD blocks\gr.

\subsection{Το Μοντέλο $\mathbf{NCDawareRank}$}

Η βασική ιδέα πίσω από τον αλγόριθμο είναι να απεικονιστεί με κάποιον τρόπο στο
τελικό μητρώο η φυσική \en NCD \gr ιδιότητα του Παγκόσμιου Ιστού. Αυτή η απεικόνιση γίνεται στο μητρώο τηλεμεταφοράς. 

Με τη χρήση παρόμοιας μαθηματικής διατύπωσης με την ενότητα $\ref{subs:pagerank}$ το αντίστοιχο μητρώο $\mathbf{G}$ διαμορφώνεται ως εξής
\begin{center}
	\begin{equation}
	\mathbf{G} = \eta \mathbf{S} + \mu \mathbf{M} + (1 - \mu - \eta) \mathbf{E}
	\label{eq:ncdawarerankmatrix}
	\end{equation}
\end{center}

Το μητρώο $ \mathbf{E}=\frac{1}{n}\boldsymbol{e} \boldsymbol{e^\top}$, όπως είναι φανερό, είναι όμοιο με το μητρώο τηλεμεταφοράς της μεθόδου \en PageRank\gr. Η παράμετρος $\mu$, με $0 > \mu > 1-\eta$ επιλέγεται έτσι ώστε να ρυθμίσει το κατά πόσο ο τυχαίος περιηγητής επιλέγει να μεταβεί από τη γραμμή διευθύνσεων σε σελίδα ενός γειτονικού \en block \gr ή σε μία οποιαδήποτε σελίδα τυχαία. Οι παράμετροι που θα χρησιμοποιήσουμε για να περιγράψουμε το μοντέλο είναι οι εξής

\begin{itemize}
\item Έστω $\mathcal{W}$ το σύνολο των κορυφών του γραφήματος και $n=|\mathcal{W}|$.
\item Συμβολίζουμε, με $u$ μια κορυφή που ανήκει στο $\mathcal{W}$, με  $\mathcal{G}_u$ το σύνολο των κορυφών που μπορούμε να επισκεφθούμε με ένα βήμα από την $u$ και με $d_u$ τον βαθμό εξόδου της $u$.
\item Θεωρούμε ένα σύνολο διαμερίσεων ${\mathcal{A}_1, \mathcal{A}_2, ... , \mathcal{A}_N}$ του $\mathcal{W}$. Κάθε τέτοιο υποσύνολο $\mathcal{A}_i$ ορίζει ένα \en  NCD block\gr. Είναι προφανές ότι για κάθε κορυφή $u \in \mathcal{W}$ υπάρχει μοναδικό $K$ τέτοιο ώστε $u \in \mathcal{A}_K$.
\item Συμβολίζουμε $\mathcal{X}_u$ το σύνολο των \en NCD blocks \gr που ανήκει μια κορυφή $u$ και οι \textit{γειτονικές} κορυφές, δηλαδή,
  \begin{center}
	$\mathcal{X}_u = \bigcup\limits_{w \in (u \cup \mathcal{G}_u)} \mathcal{A}_{(w)}$
  \end{center} 
  \item Τέλος, με $N_u$ συμβολίζουμε τον αριθμό των διαφορετικών \en NCD block \gr στο $\mathcal{X}_u$.
\end{itemize} 

Για να εξηγήσουμε την αλλαγή στη συμπεριφορά του τυχαίου περιηγητή, θεωρούμε πως το μητρώο μεταβάσεων $\mathbf{S}$ είναι μπορεί να αναλυθεί σε $N$ υποσυστήματα. Έστω πως ο τυχαίος περιηγητής βρίσκεται σε μία σελίδα $u$ που ανήκει σε ένα \en block \gr του $\mathbf{S}$. Τότε, θεωρούμε πως:
\begin{itemize}
\item με πιθανότητα $\eta$ o περιηγητής θα ακολουθήσει κάποιον από τους εξερχόμενους υπερσυνδέσμους ομοιόμορφα και με πιθανότητα $1/d_u$.
\item με πιθανότητα $1-\eta$ %$1 ? \eta$ 
θα μεταφερθεί μέσω της γραμμής διευθύνσεων σε μία νέα σελίδα και συγκεκριμένα:
\begin{itemize}
\item με πιθανότητα $\mu$ θα μεταβεί σε μία σελίδα του ίδιου \en block\gr, ή σε μια σελίδα ενός γειτονικού \en block. \gr
\item με πιθανότητα $1-\eta-\mu$ μεταβαίνει σε μία οποιαδήποτε σελίδα.
\end{itemize}\end{itemize}

Μπορούμε να εκφράσουμε το μητρώο της εξίσωσης \ref{eq:ncdawarerankmatrix} σε μορφή πιο βολική για τον υπολογισμό της δυναμομεθόδου ως εξής
\begin{center}\begin{equation} \begin{split}
\mathbf{G} &= \eta \mathbf{S} + \eta \mathbf{M} + (1-\eta - \mu) \frac{1}{n}\boldsymbol{e} \boldsymbol{e}^\top\\
&= \eta (\mathbf{H} + \frac{1}{n} \alpha \boldsymbol{e}^\top) + \mu \mathbf{M} + (1-\eta - \mu) \frac{1}{n} \boldsymbol{e} \boldsymbol{e}^\top \\
&= \eta \mathbf{H} + \mu \mathbf{M} + (\alpha \eta + (1-\eta -\mu) \boldsymbol{e})\frac{1}{n}\boldsymbol{e}^\top.
\end{split} \end{equation}\end{center}
όπου τα μητρώα $\mathbf{H}$ και $\mathbf{M}$ είναι αραιά και τα στοιχεία τους ορίζονται ως:
\begin{center}
 $\mathbf{H}_{u,v}=\begin{cases}
    \frac{1} {d_u}, & \text{αν $v \in \mathcal{G}_u$}\\
    0, & \text{αλλιώς}.
  \end{cases}$
\end{center}
και
\begin{center}
 $\mathbf{M}_{u,v}=\begin{cases}
    \frac{1} {N_u |A_{(v)}|}, & \text{αν $v \in \mathcal{X}_u$}\\
    0, & \text{αλλιώς}.
  \end{cases}$
\end{center}
Η προσθήκη του μητρώου $\mathbf{M}$ εκτός από την ποιοτική βελτίωση που προσφέρει, αφού εκφράζει πιο ρεαλιστικά τη συμπεριφορά των περιηγητών του Ιστού, βοηθάει και υπολογιστικά. Όπως είπαμε, το μητρώο \en Google\gr, $\mathbf{G}$, θεωρείται \en NCD \gr λόγω των φυσικών χαρακτηριστικών της αυτοοργάνωσης του Παγκόσμιου Ιστού. Ουσιαστικά, το μητρώο $\mathbf{M}$ συμβάλλει στο να απεικονιστεί και στο μητρώο τηλεμεταφοράς η φυσική \en NCD \gr ιδιότητα του Ιστού. 

Σύμφωνα με τα παραπάνω η δυναμομέθοδος διαμορφώνεται ως εξής
\begin{center}\begin{equation}\begin{split}
\boldsymbol{\pi^{(k+1)\top}}  & = \boldsymbol{\pi^{(k)\top}} \mathbf{G}\\
& = \eta \boldsymbol{\pi^{(k)\top}} \mathbf{S} + \mu \boldsymbol{\pi^{(k)\top}} \mathbf{M} + (1-\eta - \mu) \frac{1}{n} \boldsymbol{\pi^{(k)\top}} \boldsymbol{e} \boldsymbol{e}^\top \\
& = \eta \boldsymbol{ \pi^{(k)\top}} \mathbf{H} + \mu \boldsymbol{\pi^{(k)\top}} \mathbf{M}  +(\eta \boldsymbol{\pi^{(k)\top}} a + (1-\eta-\mu))\frac{1}{n} \boldsymbol{e}^\top.
\end{split}\end{equation}\end{center} 

Ο υπολογισμός της κατανομής πιθανοτήτων σταθερής κατάστασης γίνεται με τη χρήση της δυναμομεθόδου. Όπως είναι ήδη γνωστό από το θεώρημα $\ref{the:subdominantpagerank}$, ο ρυθμός σύγκλισης της δυναμομεθόδου όταν εφαρμόζεται σε στοχαστικά μητρώα εξαρτάται από το μέτρο της υποεπικρατούς ιδιοτιμής $\lambda_2$. Πιο συγκεκριμένα, ο ασυμπτωτικός ρυθμός σύγκλισης είναι ο ρυθμός κατά τον οποίο, $|\lambda_2 / \lambda_1|^k \rightarrow 0$. Το ακόλουθο θεώρημα διατυπώνεται στο \cite{nikolakopoulos2013ncdawarerank} και θέτει το άνω φράγμα της υποεπικρατούς ιδιοτιμής του μητρώου $\mathbf{G}$ του \en NCDawareRank\gr.

\begin{theorem}
Η υποεπικρατής ιδιοτιμή του μητρώου $\mathbf{G}$, που χρησιμοποιείται από τον \en NCDawareRank \gr αλγόριθμο και ορίζεται σύμφωνα με την εξίσωση $\ref{eq:ncdawarerankmatrix}$, έχει άνω φράγμα $\eta+\mu$.
\end{theorem}

Είναι σημαντικό να παρατηρήσουμε πως ο \en PageRank \gr θα μπορούσε να θεωρηθεί υποπερίπτωση του \en NCDawareRank \gr με δύο τρόπους:
\begin{itemize}
  \item Όταν υπάρχει ένα μοναδικό μπλοκ που περιλαμβάνει όλες τις κορυφές, το $\mathbf{Μ}$ συμπίπτει με το μητρώο τηλεμεταφοράς $\mathbf{E}$.
  \item Όταν έχουμε $K$ \en NCD blocks\gr, το καθένα από τα οποία περιλαμβάνει μια κορυφή, το $\mathbf{M}$ συμπίπτει με το μητρώο υπερσυνδέσμων $\mathbf{H}$. 
\end{itemize}
Στις δύο αυτές ακραίες περιπτώσεις, ο \en NCDawareRank \gr αλγόριθμος συμπεριφέρεται όπως ο \en PageRank\gr. Άρα θα έλεγε κανείς πως ο \en NCDawareRank \gr αποτελεί γενίκευση του \en PageRank \gr.

\subsection{Πλεονεκτήματα}

Ο αλγόριθμος  \en NCDawareRank \gr παρουσιάζει μικρότερη ευαισθησία σε προβλήματα που προκαλούνται από την αραιότητα του εκάστοτε γραφήματος στο οποίο εφαρμόζεται και αντιμετωπίζει τις νέες κορυφές, που δεν έχουν πολλές εισερχόμενες ακμές πιο δίκαια. Τέλος, ένα ακόμη σημαντικό πλεονέκτημα είναι πως μετριάζει το φαινόμενο του \en link spamming\gr.

\chapter{Κατάταξη σε διμερή γραφήματα}

Στο κεφάλαιο αυτό, θα μελετήσουμε το πρόβλημα της κατάταξης σε διμερή γραφήματα και θα προτείνουμε έναν νέο αλγόριθμο κατάταξης που θα ονομάσουμε \en BipartiteRank\gr. Ο αλγόριθμος αυτός, εκμεταλλεύεται την ιδιαίτερη \en block \gr δομή των διμερών γραφημάτων και εισάγει ένα \en block-wise \gr μητρώο τηλεμεταφοράς, διαφορετικό από αυτό του \en PageRank\gr, μιμούμενος κατά κάποιον τρόπο τον \en NCDawareRank\gr.

Αρχικά, θα δώσουμε κάποιους χρήσιμους ορισμούς  και θα περιγράψουμε την δομή του τυχαίου περιπάτου σε διμερή γραφήματα δίνοντας παραδείγματα και τις απαραίτητες αποδείξεις. Έπειτα, θα σχολιάσουμε θεωρητικά την ταχύτητα σύγκλισης του αλγορίθμου μας, η οποία είναι κατά πολύ μεγαλύτερη από αυτή του \en PageRank \gr και θα εξηγήσουμε αναλυτικά το πως προκύπτει αυτό το αποτέλεσμα.

\section{Μοντελοποίηση}

\subsection{Ορισμοί}

\begin{figure}[h!]
  \centering
      \includegraphics[width=0.5\textwidth]{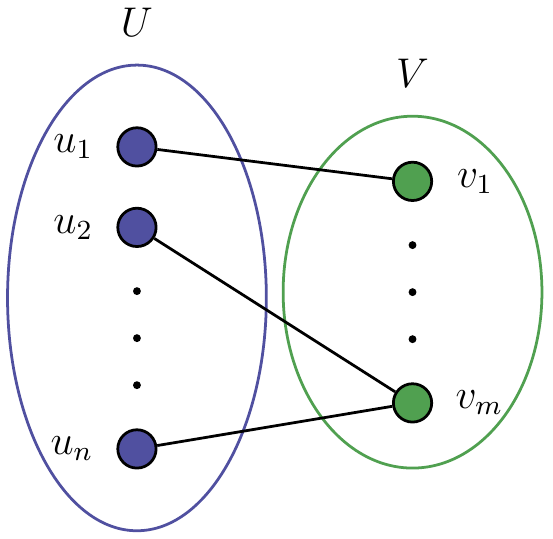}
\caption{Διμερές γράφημα} \label{fig:bipartitegraph}
\end{figure}

Ένα διμερές γράφημα είναι ένα γράφημα, του οποίου οι κορυφές μπορούν να διαιρεθούν σε δύο σύνολα $\mathcal{V}$ και $\mathcal{U}$ τέτοια ώστε να μην υπάρχει ακμή που να συνδέει μεταξύ τους δύο κορυφές του ίδιου συνόλου. Πίο τυπικά, ένα διμερές γράφημα $\mathcal{B}$ μπορεί να οριστεί ως $\mathcal{B} = (\mathcal{U} \cup \mathcal{V},\mathcal{E}_{\mathcal{B}})$, όπου $\mathcal{U}=\{u_1,u_2,...,u_m\}$ , $\mathcal{V}=\{v_1,v_2,...,v_n\}$, $\mathcal{U} \cap \mathcal{V} = \emptyset $, $\mathcal{E}_{\mathcal{B}} \subset \mathcal{V} \times \mathcal{U}$ και $\mathcal{E}_{\mathcal{B}},\mathcal{U},\mathcal{V} \neq \emptyset$. Ένα γενικό παράδειγμα φαίνεται στο σχήμα  $\ref{fig:bipartitegraph}$. Δεδομένης μιας κορυφής $i$, θα συμβολίσουμε
\begin{itemize}
\item με $N$ το πλήθος των κορυφών του $\mathcal{B}$.
\item  με $\mathcal{D}_i$ το σύνολο των γειτονικών κορυφών της $i$, και με $d_i = |\mathcal{D}_i|$ τον βαθμό εξόδου της $i$. 
\item με $\mathcal{X}_i$ το σύνολο που περιλαμβάνει την $i$, δηλαδή την $i$ και όλες τις υπόλοιπες κορυφές που ανήκουν στο ίδιο σύνολο με την $i$, και με $x_i = |\mathcal{X}_i|$ το πλήθος των κορυφών του συνόλου. 
\end{itemize}

Ας θεωρήσουμε έναν περιηγητή που ξεκινά από μια κορυφή και εκτελεί έναν τυχαίο περίπατο σε ένα διμερές γράφημα. Ο περιηγητής σε κάθε βήμα μεταβαίνει από την τρέχουσα κορυφή, σε κάποια από τις γειτονικές κορυφές με ομοιόμορφη πιθανότητα. Είναι φανερό πως οι επιλογές που έχει βρίσκονται στο αντίθετο σύνολο κάθε φορά, με συνέπεια σε άρτια βήματα να μεταβαίνει σε κορυφές του συνόλου $\mathcal{U}$ και σε περιττά σε κορυφές του συνόλου $\mathcal{V}$ ή το αντίστροφο. Την ταλάντωση αυτή υπαγορεύουν αφενός η δομή του γραφήματος και αφετέρου η περιοδικότητα του, αφού ένα διμερές γράφημα είναι περιοδικό με περίοδο 2. Όλοι οι κύκλοι του γραφήματος έχουν μήκος πολλαπλάσιο του 2.

Η ιδιαιτερότητα αυτή απεικονίζεται και στο μητρώο μεταβάσεων $\mathbf{H}$, το οποίο διαιρείται σε τέσσερα \en blocks \gr. Δύο μηδενικά τετραγωνικά \en blocks \gr στην κύρια διαγώνιο, και αυτό γιατί δεν υπάρχουν ακμές μεταξύ κορυφών του ίδιου συνόλου, και δύο μη μηδενικά ανάστροφα \en blocks \gr, τα οποία περιλαμβάνουν τις ακμές μεταξύ κορυφών που ανήκουν σε διαφορετικό σύνολο. Το $\mathbf{H}$ ορίζεται ακριβώς όπως και στον \en PageRank \gr ως:
\[ \mathbf{H}_{i,j} = \left\{
  \begin{array}{l l}
    \frac{1}{d_i} & \quad \text{αν $j \in \mathcal{D}_i$}\\
    0 & \quad \text{αλλιώς}
  \end{array} \right.\]
 
Στην ενότητα $\ref{sec:nonnegativematrices}$ αναφέραμε πως τα φαινόμενα της περιοδικότητας και της μειωσιμότητας είναι ανεπιθύμητα στις μεθόδους \en PageRank \gr και \en NCDawareRank \gr και διορθώνονται με τη βοήθεια του μητρώου τηλεμεταφοράς $\mathbf{E}$. Το μητρώο αυτό μεταφέρει τον περιηγητή σε οποιαδήποτε κορυφή του γραφήματος με ομοιόμορφη πιθανότητα. Στη μέθοδο μας, εισάγουμε ένα άλλο είδος τηλεμεταφοράς που βασίζεται στη \en block \gr δομή του γραφήματος. Ο προορισμός της τηλεμεταφοράς θεωρούμε ότι επιλέγεται ομοιόμορφα και τυχαία από όλες τις κορυφές του τρέχοντος \en block \gr κάθε φορά. Με άλλα λόγια, αν ο περιηγητής βρίσκεται στην κορυφή $i$, τότε η τηλεμεταφορά θα τον μεταφέρει σε κάθε κορυφή του $\mathcal{X}_i$ με πιθανότητα $1/x_i$. Θα μπορούσε επίσης να μεταφερθεί και στην τρέχουσα θέση του με πιθανότητα $1/x_i$.

Η ιδέα του διαφορετικού ορισμού τηλεμεταφοράς προήλθε από τον ορισμό του \en NCDawareRank\gr, ο οποίος ουσιαστικά διαφοροποιήθηκε από τα κλασικά μοντέλα τηλεμεταφοράς εκμεταλλευόμενος και την \en block \gr δομή του γραφήματος υπερσυνδέσμων μέσω του μητρώου $\mathbf{M}$. Έτσι, μας οδήγησε στην ανάπτυξη μιας παρόμοιας προσέγγισης σε διμερή γραφήματα, μιας και η \en block \gr δομή τους είναι κάτι παραπάνω από παρατηρήσιμη. Αξίζει εδώ να σημειώσουμε πως η τηλεμεταφορά στον \en NCDawareRank \gr έχει δομική αλλά και ερμηνευτική σημασία, που τεκμηριώνεται πλήρως από διάφορες μελέτες που έχουν γίνει στο παρελθόν (βλέπε ενότητα $\ref{sec:ncdawarerank}$). Ωστόσο, η προσέγγιση μας σε διμερή γραφήματα κληρονομεί μόνο τη δομική και όχι την ερμηνευτική σημασία, διότι ουσιαστικά δίνουμε τη δυνατότητα στον περιηγητή να μεταβεί σε κορυφές του ίδιου \en block\gr, ενέργεια που δεν συμπεριλαμβάνεται στον ορισμό του διμερούς γραφήματος (διότι δεν υπάρχουν μεταβάσεις μεταξύ κορυφών του ίδιου συνόλου) και συνεπώς δε θα μπορούσε να δικαιολογηθεί ερμηνευτικά. Θα κατασκευάσουμε το μητρώο  $\mathbf{M}$ του οποίου το $ij^{th}$ στοιχείο ορίζεται ως εξής
\[ \mathbf{M}_{i,j} = \left\{
  \begin{array}{l l}
    \frac{1}{x_i} & \quad \text{αν $j \in \mathcal{X}_i$}\\
    0 & \quad \text{αλλιώς}
  \end{array} \right.\]

Το μητρώο $\mathbf{M}$ έχει παρόμοια  μορφή και οργάνωση με το $\mathbf{H}$, μόνο που σε αυτή την περίπτωση τα \en block \gr της κύριας διαγωνίου περιλαμβάνουν τις ομοιόμορφες αλληλεπιδράσεις που προστίθενται μεταξύ κορυφών του ίδιου συνόλου, ενώ τα \en blocks \gr εκτός της κύριας διαγωνίου είναι μηδενικά.

Για να εκφράσουμε τη συνολική συμπεριφορά του τυχαίου περιηγητή σε αυτή τη διμερή προσέγγιση θα κατασκευάσουμε ένα νέο μητρώο:
\begin{equation} \label{eq:bincdawarerankmatrix}
	\mathbf{P} = \epsilon \mathbf{H} + (1-\epsilon) \mathbf{M}.
\end{equation}
Θα μπορούσαμε αντί για το $\mathbf{Η}$ να χρησιμοποιήσουμε το $\mathbf{S}$ το οποίο προκύπτει μετά από προσαρμογή στοχαστικότητας στο $\mathbf{H}$ (βλέπε ενότητα $\ref{RandomSurferModel}$), ωστόσο αφού το γράφημα είναι μη κατευθυνόμενο, μπορούν να προκύψουν μηδενικές γραμμές μόνο σε περίπτωση που υπάρχουν εκκρεμείς κορυφές. Θεωρούμε πως δεν έχει νόημα να συμπεριλάβουμε αυτές τις κορυφές στο μοντέλο. Επιπλέον, το μητρώο $\mathbf{E}$  δεν συμμετέχει ούτε κατά ένα μικρό ποσοστό στη διαμόρφωση της τελικής συμπεριφοράς του περιηγητή όπως στους αλγορίθμους \en PageRank \gr και \en NCDawareRank\gr. Όπως θα αποδείξουμε στη συνέχεια, η  συμμετοχή του μητρώου $\mathbf{M}$ στη διαμόρφωση του τελικού μητρώου $\mathbf{P}$ εξασφαλίζει ότι το τελευταίο είναι πρωταρχικό. 

\begin{theorem} \label{the:ergodicityofP}
Η αλυσίδα \en Markov \gr πεπερασμένων καταστάσεων  με μητρώο πιθανοτήτων μετάβασης το $\mathbf{P}$, είναι αμείωτη και μη περιοδική (εργοδική).
\end{theorem}

\begin{proof}

\begin{figure} 
  \centering
      \includegraphics[width=0.7\textwidth]{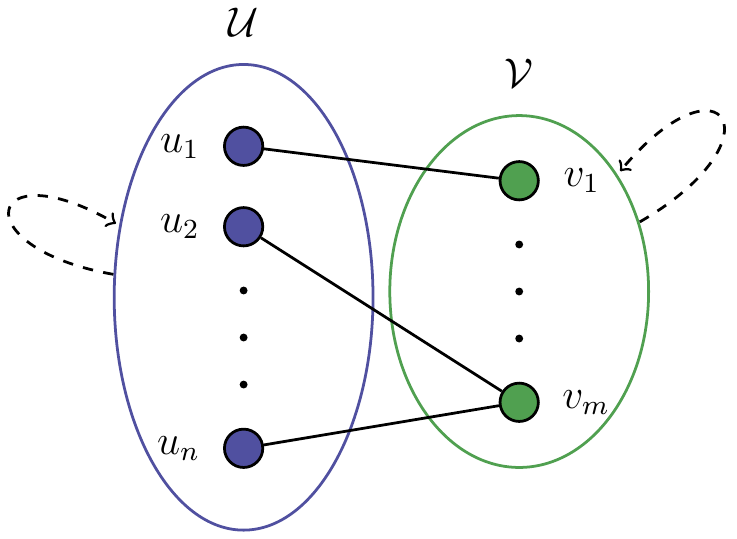}
\caption{Προσαρμογή του τυχαίου περιπάτου στο διμερές γράφημα του σχήματος \ref{fig:bipartitegraph} προσθέτοντας μεταβάσεις (διακεκομμένα βέλη) μεταξύ κορυφών που ανήκουν στο ίδιο σύνολο.} \label{fig:bipartiterandomwalk}
\end{figure}

Θεωρούμε ένα γράφημα  $\mathcal{G}=(\mathcal{U} \cup \mathcal{V}, \mathcal{E} )$, όπου $\mathcal{U} \cap \mathcal{V} = \emptyset$ και $\mathcal{U},\mathcal{V} \neq \emptyset$. Τα υπογραφήματα $\mathcal{G}_\mathcal{U} = (\mathcal{U}, \mathcal{E}_\mathcal{U})$ και $\mathcal{G}_\mathcal{V}=(\mathcal{V}, \mathcal{E}_\mathcal{V})$ είναι πλήρη γραφήματα με επιπλέον ιδιοβρόγχους σε κάθε κορυφή και προέκυψαν από τις επιπρόσθετες ακμές που προστέθηκαν στο συνολικό γράφημα μέσω του μητρώου $\mathbf{M}$. Το $\mathcal{B}=(\mathcal{V} \times \mathcal{U},\mathcal{E}_\mathcal{B})$ είναι ένα διμερές γράφημα που εκφράζεται από το μητρώο $\mathbf{H}$. Ορίζουμε το σύνολο των κορυφών $\mathcal{E}$ ως $\mathcal{E} = (\mathcal{E}_\mathcal{U} \cup \mathcal{E}_\mathcal{V} \cup \mathcal{E}_\mathcal{B})$ με $\mathcal{E}_\mathcal{B} \neq \emptyset$. Το γράφημα που μόλις περιγράψαμε φαίνεται στο σχήμα \ref{fig:bipartiterandomwalk}.

\begin{lemma} \label{lemma:connectedgraph}
Μια αλυσίδα \en Markov \gr πεπερασμένων καταστάσεων  είναι αμείωτη αν το γράφημα που την αναπαριστά είναι ένα ισχυρά συνδεδεμένο γράφημα. 
\end{lemma}

\begin{lemma} \label{lemma:connectedgraph2}
Ένα μη κατευθυνόμενο γράφημα είναι ισχυρά συνδεδεμένο όταν δεν υπάρχει διαμέριση των κορυφών του σε δύο σύνολα $\mathcal{I}$ και $\mathcal{J}$, τέτοια ώστε να μην υπάρχει καμία ακμή μεταξύ των δύο συνόλων.
\end{lemma}

Έστω ότι το γράφημα $\mathcal{G}$ δεν είναι ισχυρά συνδεδεμένο. Θα πρέπει να αποδείξουμε πως υπάρχει διαμέριση του $G=(\mathcal{I},\mathcal{J})$ όπου $\mathcal{I} \cap \mathcal{J} = \emptyset$, ώστε να μην υπάρχει ακμή που να ενώνει δύο κορυφές $i$ και $j$, όπου $i \in \mathcal{I}$ και $j \in \mathcal{J}$ (λήμματα $\ref{lemma:connectedgraph}$ και $\ref{lemma:connectedgraph2}$). Διακρίνουμε τις εξής περιπτώσεις:
\begin{enumerate}
\item Έστω $\mathcal{I} = \mathcal{U}$ και $\mathcal{J} = \mathcal{V}$. Τότε υπάρχει τουλάχιστον μια ακμή μεταξύ $\mathcal{I}$ και $\mathcal{J}$, αφού $\mathcal{E} = (\mathcal{E}_\mathcal{U} \cup \mathcal{E}_\mathcal{V} \cup \mathcal{E}_\mathcal{B})$ με $\mathcal{E}_\mathcal{B} \neq \emptyset$ από τον ορισμό των γραφημάτων $\mathcal{G}$ και $\mathcal{B}$.

\item Έστω $\mathcal{I}=\{v\}$ όπου $v \in \mathcal{V}$. 
Συνεπώς, $\mathcal{J} = \mathcal{U} \cup \mathcal{V} - \{v\}$. Τότε, υπάρχει τουλάχιστον μία ακμή μεταξύ των συνόλων $\mathcal{I}$ και $\mathcal{J}$, αφού το $\mathcal{G_V}$ είναι ένα πλήρες γράφημα.  Από αυτό συνεπάγεται πως υπάρχουν ακμές που συνδέουν την $v$ με όλες της κορυφές του
$\mathcal{V}-\{v\}$ οι οποίες ανήκουν στο σύνολο $\mathcal{J}$. Σημειώνουμε πως η περίπτωση που η κορυφή $v$ είναι η μοναδική κορυφή του συνόλου $\mathcal{V}$ έχει ήδη ελεγχθεί στο 1.

\item Έστω $\mathcal{I}=\{u\}$ όπου $u \in \mathcal{U}$. 
Συνεπώς, $\mathcal{J} = \mathcal{V} \cup \mathcal{U} - \{u\}$. 
Τότε, υπάρχει τουλάχιστον μία ακμή μεταξύ των συνόλων $\mathcal{I}$ και $\mathcal{J}$, αφού το $\mathcal{G_U}$ είναι ένα πλήρες γράφημα. Από αυτό συνεπάγεται πως υπάρχουν ακμές που συνδέουν την $u$ με όλες της κορυφές του
$\mathcal{U}-\{u\}$, οι οποίες ανήκουν στο σύνολο $\mathcal{J}$. Σημειώνουμε πως η περίπτωση που η κορυφή $u$ είναι η μοναδική κορυφή του συνόλου $\mathcal{U}$ έχει ήδη ελεγχθεί στο 1.
\end{enumerate}

Είναι προφανές ότι οποιαδήποτε άλλη περίπτωση οδηγεί επίσης σε άτοπο. Άρα το γράφημα $\mathcal{G}$ είναι ισχυρά συνδεδεμένο και η αντίστοιχη αλυσίδα \en Markov \gr που περιγράφει τον τυχαίο περίπατο σε ένα διμερές γράφημα είναι αμείωτη.

\begin{lemma}
Σε μια αμείωτη αλυσίδα \en Markov\gr, όλες οι καταστάσεις έχουν την ίδια περίοδο.
\end{lemma}

\begin{lemma} \label{lemma:periodicchain}
Μια αμείωτη αλυσίδα \en Markov \gr είναι μη περιοδική αν τουλάχιστον μια από τις καταστάσεις της είναι μη περιοδική.
\end{lemma}

Από τον ορισμό του γραφήματος $\mathcal{G}$, λόγω των ιδιοβρόγχων που υπάρχουν στα $\mathcal{G_U}=(\mathcal{U},\mathcal{E}_\mathcal{U})$ και $\mathcal{G_V}=(\mathcal{V},\mathcal{E}_\mathcal{V})$ υπάρχει τουλάχιστον μια κορυφή με περιοδικότητα 1 (μη περιοδική κορυφή). Άρα το γράφημα αναπαριστά μια μη περιοδική αλυσίδα \en Markov \gr (λήμμα $\ref{lemma:periodicchain}$).

Τέλος, αφού αναφερόμαστε σε αλυσίδες \en Markov \gr πεπερασμένων καταστάσεων, η μη μειωσιμότητα και μη περιοδικότητα εξασφαλίζουν πως η αλυσίδα είναι εργοδική, δηλαδή σύμφωνα με το θεώρημα $\ref{the:ergodicity}$ συγκλίνει σε μια μοναδική σταθερή κατανομή πιθανοτήτων.
\end{proof}

Σύμφωνα με το πρώτο σκέλος του θεωρήματος $\ref{the:primitivity}$ ένα μητρώο μη αρνητικό, αμείωτο που έχει μια ιδιοτιμή πάνω στον φασματικό κύκλο (μη περιοδικό) είναι πρωταρχικό. Το $\mathbf{P}$ πληρεί τις προϋποθέσεις αυτές αφού είναι στοχαστικό, αμείωτο και μη περιοδικό $\ref{the:ergodicityofP}$. Συνεπώς το $\mathbf{P}$ είναι ένα πρωταρχικό μητρώο.

\subsubsection{Αποθήκευση Μητρώων}

Το μητρώο μεταβάσεων έχει μια \en block \gr μορφή και γράφεται 
\begin{center} \[ \mathbf{H}=
\begin{bmatrix}
	\mathbf{0} & \mathbf{L} \\ \mathbf{L}^\intercal & \mathbf{0}
\end{bmatrix}
\] \end{center} 
όπου με $\mathbf{L}$ συμβολίζουμε το μητρώο μεταβάσεων από το σύνολο $\mathcal{U}$ στο $\mathcal{V}$ και με $\mathbf{L}^\intercal$ το μητρώο μεταβάσεων από το σύνολο $\mathcal{V}$ στο $\mathcal{U}$ αντίστοιχα. Το $\mathbf{H}$ μπορεί να περιγραφεί πλήρως από το $\mathbf{L}$, του οποίου ο αριθμός των μη αρνητικών στοιχείων φράσσεται από το $n \cdot m$.

Επιπλέον, το μητρώο $\mathbf{M}$ δεν χρειάζεται να αποθηκευτεί καθόλου αφού οι τιμές του εξαρτώνται μόνο από τα μεγέθη των \en blocks\gr. Μπορεί όμως πρακτικά να εκφραστεί και ως γινόμενο δύο αραιών μητρώων (όπως στο  \cite{nikolakopoulos2013ncdawarerank}), δηλαδή $\mathbf{M} = \mathbf{R} \mathbf{A}$. 
Το μητρώο $\mathbf{A} \in \mathbb{R}^{2 \times N}$ ορίζεται ως:
\begin{center} \[ \mathbf{A}=
\begin{bmatrix}
	 \boldsymbol{e}^\intercal_{n} & 0           \\
       0 &   \boldsymbol{e}^\intercal_{m}
\end{bmatrix}
\] \end{center} 
όπου  $\boldsymbol{e}^\intercal_{\left|\mathcal{K}\right|}$ εκφράζει ένα διάνυσμα στο $\mathbb{R}^\mathcal{K}$
του οποίου τα στοιχεία είναι όλα 1, ενώ το μητρώο $\mathbf{R} \in \mathbb{R}^{N \times 2}$ ορίζεται ως:
\begin{center} \[ \mathbf{R}=
\begin{bmatrix}
	 \boldsymbol{e}_{n}/n & 0           \\
       0 &   \boldsymbol{e}_{m}/m
\end{bmatrix}
\] \end{center} 
Τα μητρώα $\mathbf{R}$ και $\mathbf{A}$ έχουν μόνο $N$ μη μηδενικά στοιχεία το καθένα.

Υπάρχουν συμπαγείς μηχανισμοί αποθήκευσης που εκμεταλλεύονται την αραιότητα των μητρώων που μόλις αναφέραμε.

\subsubsection{Παράδειγμα}

\begin{figure}
\centering
\includegraphics[width=0.3\textwidth]{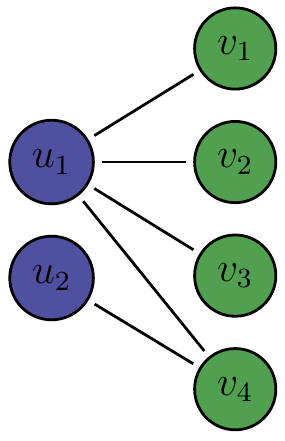}
\caption{Παράδειγμα διμερούς γραφήματος.}
\label{fig:exampleGraph}
\end{figure}

Αφού δώσαμε τους ορισμούς των μητρώων που συμμετέχουν στον υπολογισμό του διανύσματος κατάταξης της μεθόδου μας, θα δημιουργήσουμε ένα απλό παράδειγμα ώστε να γίνει κατανοητός ο τρόπος κατασκευής τους. Θεωρούμε το διμερές γράφημα του σχήματος \ref{fig:exampleGraph}. Το γράφημα αποτελείται από δύο σύνολα κορυφών τα οποία αποτελούνται από δύο και τέσσερις κορυφές αντίστοιχα. Το μητρώο μεταβάσεων $\mathbf{H}$ που αντιστοιχεί σε αυτό το γράφημα είναι:

\begin{center}  \[ \mathbf{H}=
\begin{bmatrix}
    0  &  0  &  \textcolor{red}{1/4}  &  \textcolor{red}{1/4} & \textcolor{red}{1/4} & \textcolor{red}{1/4}\\
    0  &  0  &  \textcolor{red}{0}  &  \textcolor{red}{0} & \textcolor{red}{0} & \textcolor{red}{1}\\
    \textcolor{red}{1}  &  \textcolor{red}{0}  &  0  &  0 & 0 & 0\\
    \textcolor{red}{1}  & \textcolor{red}{0}  &  0  &  0 & 0 & 0\\
    \textcolor{red}{1}  &  \textcolor{red}{0}  &  0  &  0 & 0 & 0\\
    \textcolor{red}{1/2}  &  \textcolor{red}{1/2}  &  0  &  0 & 0 & 0\\
\end{bmatrix}
\]  \end{center}
Το μητρώο $\mathbf{M}$ αντίστοιχα θα είναι:
\begin{center} \[ \mathbf{M}=
\begin{bmatrix}
    \textcolor{red}{1/2}  &  \textcolor{red}{1/2}  &  0  &  0 & 0 & 0\\
    \textcolor{red}{1/2}  &  \textcolor{red}{1/2}  &  0  &  0 & 0 & 0\\
    0  &  0  &  \textcolor{red}{1/4}  &  \textcolor{red}{1/4} & \textcolor{red}{1/4} & \textcolor{red}{1/4}\\
    0  &  0  &  \textcolor{red}{1/4}  &  \textcolor{red}{1/4} & \textcolor{red}{1/4} & \textcolor{red}{1/4}\\
    0  &  0  &  \textcolor{red}{1/4}  &   \textcolor{red}{1/4} &  \textcolor{red}{1/4} &  \textcolor{red}{1/4}\\
    0  &  0  &   \textcolor{red}{1/4}  &   \textcolor{red}{1/4} &  \textcolor{red}{1/4} &  \textcolor{red}{1/4}\\
\end{bmatrix}
\] \end{center}
και μπορεί να εκφραστεί ως γινόμενο $\mathbf{R} \mathbf{A}$ με:
\begin{center} \[ \mathbf{R}=
\begin{bmatrix}
     \textcolor{red}{1/2} &  0 \\
     \textcolor{red}{1/2}  &  0 \\
    0  &   \textcolor{red}{1/4} \\
    0  &   \textcolor{red}{1/4}  \\
    0  &   \textcolor{red}{1/4}  \\
    0  &   \textcolor{red}{1/4}  \\
\end{bmatrix}
\] \end{center}
και
\begin{center} \[ \mathbf{A}=
\begin{bmatrix}
     \textcolor{red}{1} &  \textcolor{red}{1} & 0 & 0 & 0 & 0\\
    0 &  0 & \textcolor{red}{1} & \textcolor{red}{1} & \textcolor{red}{1} & \textcolor{red}{1}\\
\end{bmatrix}
\] \end{center}

\subsection{Τυχαίος Περίπατος και $\mathbf{block-wise}$ Τηλεμεταφορά}

Συνοψίζοντας, περιηγητής στην μέθοδό μας χρησιμοποιεί την πράξη της τηλεμεταφοράς με δύο τρόπους:
\begin{itemize}
\item όταν βρεθεί σε μια κορυφή χωρίς εξερχόμενες ακμές,
\item όταν βρεθεί σε κορυφή με εξερχόμενες ακμές ενεργοποιεί την τηλεμεταφορά με πιθανότητα  $1 - \epsilon$ ή συνεχίζει τον τυχαίο περίπατο με πιθανότητα $\epsilon$, με $0< \epsilon <1$. Κατά την τηλεμεταφορά, ο περιηγητής θα μεταβεί σε μια κορυφή του συνόλου στο οποίο βρίσκεται (συμπεριλαμβανομένης και της τρέχουσας) τυχαία και με ομοιόμορφη πιθανότητα.
\end{itemize}

\subsection{Δυναμομέθοδος}

Αν η δυναμομέθοδος εφαρμοστεί στο $\mathbf{P}$ μπορεί να γραφεί:
\begin{center}\begin{equation}\label{eq:powermethodbipartiterank} \begin{split}
\boldsymbol{\pi^{(k+1)\top}}  & = \boldsymbol{\pi^{(k)\top}} \mathbf{P}\\
& = \epsilon \boldsymbol{\pi^{(k)\top}} \mathbf{H} + (1-\epsilon) \boldsymbol{\pi^{(k)\top}} \mathbf{R} \mathbf{A}.
\end{split}\end{equation}\end{center} 
Όπως αναφέραμε και προηγουμένως, τα $\mathbf{R}$ και $\mathbf{A}$ δεν χρειάζεται να αποθηκευτούν αφού το κάθε στοιχείο του διανύσματος του γινομένου $\boldsymbol{\pi^{(k)\top}} \mathbf{R} \mathbf{A}$ μπορεί να εκφραστεί ως
\begin{equation*}
\frac{1}{x_i} \sum \limits_{j=1}^{x_i} \boldsymbol{\pi_{j}}
\end{equation*}
όπου το $x_i$ είναι το πλήθος των κορυφών του συνόλου $i$. Οι πράξεις \en MV \gr αφορούν και εδώ το ιδιαίτερα αραιό μητρώο $\mathbf{H}$.

\begin{algorithm}
\caption{$\mathbf{BipartiteRank}$}
\label{CHalgorithm}
\begin{algorithmic}[1]
\\ Οργανώνουμε τις κορυφές του μητρώου $\mathbf{H}$ σύμφωνα με τη διαμέριση {$\mathcal{U,V}$}.
\\ Κατασκευάζουμε τα μητρώα $\mathbf{R}$ και $\mathbf{A}$ σύμφωνα με τη διαμέριση  {$\mathcal{U,V}$}
\\ Λύνουμε το ακόλουθο πρόβλημα ιδιοτιμών: \begin{center} $\pi^\top = \pi^\top \mathbf{P}$ \\ $\pi^\top \boldsymbol{e} = 1$ \end{center}
χρησιμοποιώντας το μητρώο $\mathbf{P}$ εκφρασμένο με χρήση της δυναμομεθόδου, όπως στην εξίσωση $\ref{eq:powermethodbipartiterank}$.
\end{algorithmic}
\end{algorithm}

\subsubsection{Παράδειγμα}

Θεωρούμε το γράφημα του σχήματος $\ref{fig:exampleGraph}$, με $\epsilon = 0.85$. Τότε το μητρώο μεταβάσεων του περιηγητή με τηλεμεταφορά είναι:
\begin{center} \[ \mathbf{P}=
\begin{bmatrix}
     0.0750  &  0.0750 &   0.2125  &  0.2125  &  0.2125  &  0.2125\\
	 0.0750  &  0.0750 &   0  &  0  &  0  &  0.8500\\
 	 0.8500  &  0 &   0.0375  &  0.0375  &  0.0375  &  0.0375\\
	 0.8500  &  0 &   0.0375  &  0.0375  &  0.0375  &  0.0375\\
	  0.8500  &  0 &   0.0375  &  0.0375  &  0.0375  &  0.0375\\
	 0.4250  &  0.4250 &   0.0375  &  0.0375  &  0.0375  &  0.0375\\
\end{bmatrix}
\] \end{center}
Έστω ότι ο περιηγητής ξεκινά από την κατάσταση 1, η οποία αντιστοιχεί στο αρχικό διάνυσμα κατανομής πιθανοτήτων:
\begin{center} \[ \boldsymbol{\pi^{(1)\top}}=
\begin{bmatrix}
     0.1667 & 0.1667 & 0.1667 & 0.1667 & 0.1667 & 0.1667\\
\end{bmatrix}
\] \end{center}
Μετά από ένα βήμα η κατανομή θα είναι:
\begin{center}\[ \boldsymbol{\pi^{(1)\top}}\mathbf{P}= 
\begin{bmatrix}
0.0750 &   0.0750 &   0.2125  &  0.2125 &   0.2125  &  0.2125\\
\end{bmatrix} = \boldsymbol{\pi^{(2)\top}}.
\] \end{center}
Αν συνεχίσουμε για πολλά ακόμα βήματα θα παρατηρήσουμε ότι η κατανομή συγκλίνει στη σταθερή κατάσταση:
\begin{center}\[ \boldsymbol{\pi^{(k-1)\top}}\mathbf{P}= 
\begin{bmatrix}
0.3757  &  0.1243  &  0.0986  &  0.0986 &   0.0986  &  0.2042\\
\end{bmatrix} = \boldsymbol{\pi^{(k)\top}}.
\] \end{center}

\section{Ταχύτητα Σύγκλισης} \label{subs:rateofconvergence}

Σύμφωνα με το θεώρημα $\ref{the:subdominantpagerank}$, ο ρυθμός σύγκλισης της δυναμομεθόδου όταν εφαρμόζεται σε στοχαστικά μητρώα εξαρτάται από την απόλυτη τιμή της υποεπικρατούς ιδιοτιμής $\lambda_2$ και είναι ο ρυθμός κατά τον οποίο, $|\lambda_2 / \lambda_1|^k \rightarrow 0$. Αποδείχθηκε πως το τελικό μητρώο $\mathbf{P}$ είναι πρωταρχικό και συνεπώς, το φάσμα του συνθέτουν πραγματικές ιδιοτιμές εκ των οποίων η μία μόνο βρίσκεται πάνω στον φασματικό κύκλο (βλέπε θεώρημα $\ref{the:primitivity}$). Η ιδιοτιμή αυτή έχει την τιμή 1 λόγω του ότι το μητρώο είναι στοχαστικό. Μια σημαντική παρατήρηση είναι ότι το μητρώο $\mathbf{P}$, οποιοδήποτε κι αν είναι, έχει μια ιδιοτιμή η οποία εξαρτάται μόνο από τον παράγοντα τηλεμεταφοράς $\epsilon$. Πιο συγκεκριμένα,

\begin{theorem}
Το μητρώο $\mathbf{P}$ που ορίζεται από την εξίσωση $\ref{eq:ncdawarerankmatrix}$ έχει μια ιδιοτιμή ίση με $(1-\epsilon) - \epsilon$.
\end{theorem}

\begin{proof}
Παρατηρώντας το τελικό μητρώο $\mathbf{P}$, είναι φανερό πως έχει ένα ιδιοδιάνυσμα \begin{center}
$\boldsymbol{\upsilon} = [1,1,...,1,-1,-1,...,-1]$
\end{center} του οποίου το πλήθος των $1$ είναι  $n$ και το πλήθος των $-1$ είναι $m$. Υπενθυμίζουμε πως ιδιοδιάνυσμα ενός τετραγωνικού μητρώου $\mathbf{P}$, είναι ένα μη μηδενικό διάνυσμα $\boldsymbol{\upsilon}$ που όταν πολλαπλασιαστεί με τον $\mathbf{P}$, ισούται με το αρχικό διάνυσμα, πολλαπλασιασμένο με έναν αριθμό $\lambda$, έτσι ώστε: $\mathbf{P} \boldsymbol{\upsilon} = \lambda \boldsymbol{\upsilon}$. Ο αριθμός $\lambda$ ονομάζεται ιδιοτιμή του $\mathbf{P}$ που αντιστοιχεί στο $\boldsymbol{\upsilon}$.

Άρα έχουμε

\begin{center}{\begin{equation} \label{eq:proofeigenvalue}
\begin{split}
\mathbf{P} \boldsymbol{\upsilon} & = \epsilon \mathbf{H} \boldsymbol{\upsilon} + (1-\epsilon) \mathbf{M}\boldsymbol{\upsilon} \\
& = \epsilon \begin{bmatrix}
     & & & \mathbf{H}_{1,n+1} & ... & \mathbf{H}_{1,n+m} \\
     & \mathbf{0} &  & \vdots & \ddots & \vdots \\
     & & & \mathbf{H}_{m,n+1} & ... & \mathbf{H}_{m,n+m} \\
     \mathbf{H}_{n+1,1} & ... & \mathbf{H}_{n+1,n} &  & & \\
     \vdots & \ddots & \vdots &  & \mathbf{0} & \\
     \mathbf{H}_{n+m,1} & ... & \mathbf{H}_{n+m,n} &  & & \\
\end{bmatrix} \begin{bmatrix}
      1\\
      \vdots\\
      1\\
      -1\\
      \vdots\\
      -1\\
\end{bmatrix}  \\
&+ (1-\epsilon) \begin{bmatrix}
     \mathbf{M}_{1,1} & \cdots & \mathbf{M}_{1,n} & & &  \\
     & \vdots & \ddots &  & \mathbf{0} &   \\
      \mathbf{M}_{n,1} & \cdots &  \mathbf{M}_{n,n}&  &  &\\
      & & &  \mathbf{M}_{n+1,n+1} & \cdots &  \mathbf{M}_{n+1,m}\\
      & \mathbf{0} & & \vdots &\ddots & \vdots\\
      &  &  & \mathbf{M}_{m,n+1} &\cdots & \mathbf{M}_{m,m} \\
\end{bmatrix} \begin{bmatrix}
      1\\
      \vdots\\
      1\\
      -1\\
      \vdots\\
      -1\\
\end{bmatrix}
\end{split}\end{equation}}\end{center} 

Λόγω της στοχαστικότητας των μητρώων $\mathbf{H}$ και $\mathbf{M}$ η σχέση της εξίσωσης \ref{eq:proofeigenvalue} είναι ίση με 
\begin{center}
$\epsilon \begin{bmatrix}
      -1\\
      \vdots\\
      -1\\
      1\\
      \vdots\\
      1\\
\end{bmatrix} + (1-\epsilon)\begin{bmatrix}
      1\\
      \vdots\\
      1\\
      -1\\
      \vdots\\
      -1\\
\end{bmatrix} $
\end{center}

Άρα έχουμε \begin{center}
\begin{equation} \begin{split}
\mathbf{P} \boldsymbol{\upsilon}  &=-\epsilon \begin{bmatrix}
      1\\
      \vdots\\
      1\\
      -1\\
      \vdots\\
      -1\\
\end{bmatrix} + (1-\epsilon)\begin{bmatrix}
      1\\
      \vdots\\
      1\\
      -1\\
      \vdots\\
      -1\\
\end{bmatrix}\\
&= ((1-\epsilon) - \epsilon) \begin{bmatrix}
      1\\
      \vdots\\
      1\\
      -1\\
      \vdots\\
      -1\\
\end{bmatrix}\\
&-= ((1-\epsilon) - \epsilon) \boldsymbol{\upsilon}
\end{split}\end{equation} 
\end{center}
\end{proof}

Συνεπώς, η τιμή  $(1-\epsilon) - \epsilon$ είναι ιδιοτιμή του $\mathbf{P}$ και αντιστοιχεί στο ιδιοδιάνυσμα $\boldsymbol{\upsilon}$. Στην πλειοψηφία των περιπτώσεων, και ειδικότερα όταν στην κατασκευή του $\mathbf{P}$ εμπλέκονται γραφήματα του πραγματικού κόσμου, η ιδιοτιμή αυτή είναι η υποεπικρατής ιδιοτιμή του $\mathbf{P}$, δηλαδή $|\lambda_2|= |(1-\epsilon) - \epsilon|$. Αυτό είναι μια σπουδαία παρατήρηση, αν σκεφτεί κανείς πως η σύγκλιση σε αυτές της περιπτώσεις εξαρτάται μόνο από την παράμετρο $\epsilon$. 

Ακόμη, το γεγονός ότι το $\mathbf{H}$ είναι περιοδικό με 2 περίοδο σημαίνει ότι έχει 2 ιδιοτιμές πάνω στον φασματικό του κύκλο ίσες με 1 κ -1, άρα για την υποεπικρατής ιδιοτιμη του τελικού μητρώου του \en PageRank \gr θα ισχύει $|\lambda_2|= \epsilon$ (βλέπε το θεώρημα $\ref{the:subdominantpagerank}$). Συνεπώς, ο αλγόριθμος \en BipartiteRank \gr αναμένεται να συγκλίνει σε λιγότερα από τον \en PageRank\gr, γεγονός που θα δείξουμε και πειραματικά στη συνέχεια.

\chapter{Πειραματική Αξιολόγηση}

Στο κεφάλαιο αυτό θα αξιολογήσουμε την μέθοδο μας κάνοντας κατάταξη σε πραγματικά δεδομένα. Αρχικά θα παραθέσουμε τους κώδικες των αλγορίθμων σε \en Matlab \gr και σε \en C\gr. Έπειτα θα περιγράψουμε τα δεδομένα μας και θα αναφέρουμε τις πηγές από τις οποίες αντλήθηκαν. Τέλος, θα συγκρίνουμε πειραματικά τις υπολογιστικές επιδόσεις του αλγορίθμου \en BipartiteRank \gr με τις αντίστοιχες του αλγορίθμου \en PageRank\gr.

\section{Υλοποίηση}

Υλοποιήσαμε τον αλγόριθμο \en BipartiteRank \gr πρώτα σε \en Matlab \gr και έπειτα σε \en C \gr με τη βοήθεια της εφαρμογής \en MATLAB Coder\gr. 
%Ο κώδικας σε \en C \gr, λόγω της μεγάλης έκτασης του παρατίθεται στο παράρτημα.

%\lstinputlisting{../codes/BipartiteRank.m} 
\gr
\section{Δεδομένα}

Χρησιμοποιήσαμε έξι σύνολα πραγματικών δεδομένων \cite{kunegis2013konect} των οποίων τα χαρακτηριστικά φαίνονται στον πίνακα \ref{table:datasets}. 

\begin{table}[h!]
\caption{Δεδομένα}
\begin{center}
\scalebox{0.75}{
\begin{tabular}{l*{7}{c}  r }
\hline
Δεδομένα              & Γραμμές & Στήλες & Μη μηδενικά & Πυκνότητα & Συνδεδεμένο\\
\hline 
\en Jester\gr 			  & $73.421$ & $100$ & $4.136.360$ & $0.5634$ &  Ναι\\
\en MovieLens10M \gr      & $69.878$ & $10.677$ & $10.000.054$ & $0.0143$ & Ναι \\
\en TREC\gr     		& $551.787$ & $1.173.225$ & $83.629.405$ & $1.2918  \cdot 10^{-4}$ & Ναι \\
\en Reuters\gr         & $781.265$ & $283.911$ & $60.569.726$ & $2.7307  \cdot 10^{-4}$ & Ναι \\
\en DBLP\gr     		& $1.425.813$ & $4.000.150$ & $8.649.016$ & $1.5164  \cdot 10^{-6}$ & Όχι \\
\en YoutubeMemberships\gr & $94.238$ & $124.325$ & $293.360$ & $1.0347  \cdot 10^{-4}$ & Όχι\\ \hline 
\end{tabular} \label{table:datasets}
}
\end{center}
\end{table}

\begin{itemize}
\item \en Jester\gr: Περιλαμβάνει τις βαθμολογίες χρηστών σε 100 αστεία. Τα σύνολα κορυφών αντιστοιχούν σε χρήστες και αστεία. Οι ακμές αντιπροσωπεύουν βαθμολογίες οι οποίες κυμαίνονται από -10 έως 10. Οι βαθμολογίες έχουν κανονικοποιηθεί ώστε να είναι θετικές.\\
\item \en MovieLens10M\gr: Περιλαμβάνει δέκα εκατομμύρια βαθμολογίες ταινιών. Τα σύνολα κορυφών αντιστοιχούν σε χρήστες και ταινίες.  Οι ακμές αντιπροσωπεύουν και σε αυτήν την περίπτωση βαθμολογίες χρηστών.\\
\item \en TREC\gr: Προέρχεται από αρχεία κειμένου του \en TREC\gr. Τα σύνολα κορυφών αντιστοιχούν σε κείμενα και λέξεις. Κάθε βεβαρημένη ακμή υποδηλώνει την συχνότητα με την οποία εμφανίζεται μια λέξη σε ένα κείμενο.\\
\item \en Reuters\gr: Περιλαμβάνει αναφορές λέξεων σε κείμενα που προέρχονται από το ειδησεογραφικό πρακτορείο \en Reuters (RCV1)\gr. Τα σύνολα αντιστοιχούν σε κείμενα και λέξεις. Κάθε βεβαρημένη ακμή υποδηλώνει την συχνότητα με την οποία εμφανίζεται μια λέξη σε ένα κείμενο. \\
\item \en DBLP \gr: Προέρχεται από την επιστημονική βιβλιογραφία και τα αρχεία του \en DBLP\gr. Τα σύνολα αντιστοιχούν σε συγγραφείς και δημοσιεύσεις. Οι ακμές ενώνουν κάθε συγγραφέα με τις δημοσιεύσεις του. Λόγω του ότι είναι μη συνδεδεμένο, για την πραγματοποίηση των υπολογιστικών πειραμάτων, κάναμε \en cleaning \gr κρατώντας μόνο το μεγαλύτερο συνδεδεμένο τμήμα του (βλέπε ενότητα $\ref{sub:disconnectedcomponents}$). Μετά το \en cleaning \gr το γράφημα περιλαμβάνει 1,210,591 συγγραφείς, 3,605,818 δημοσιεύσεις και οι ακμές του είναι 8,142,407.\\
\item \en YoutubeMemberships\gr: Περιλαμβάνει χρήστες του \en Youtube \gr και τη συμμετοχή τους σε ομάδες.  Το ένα σύνολο κορυφών αντιστοιχεί σε χρήστες και το άλλο σε ομάδες. Κάθε ακμή αντιπροσωπεύει την συμμετοχή ενός χρήστη σε ομάδα. Λόγω του ότι είναι μη συνδεδεμένο, για την πραγματοποίηση των υπολογιστικών πειραμάτων, κάναμε \en cleaning \gr κρατώντας μόνο το μεγαλύτερο συνδεδεμένο τμήμα του (βλέπε ενότητα $\ref{sub:disconnectedcomponents}$). Μετά το \en cleaning \gr το γράφημα περιλαμβάνει 88,490 χρήστες, 25,007 ομάδες χρηστών και οι ακμές του είναι 573,826. \\
\end{itemize}

%Οι κώδικες για την κατασκευή των απαραίτητων μητρώων είναι οι εξής\\
%\\
%\en \textbf{ReadDataset.m} 
%\lstinputlisting{../codes/ReadDataset.m}
%\textbf{ReadCleanDataset.m} 
%\lstinputlisting{../codes/ReadCleanDataset.m} \gr

\section{Υπολογιστικά Πειράματα}

\begin{figure}[h!] 
\includegraphics[width=\textwidth]{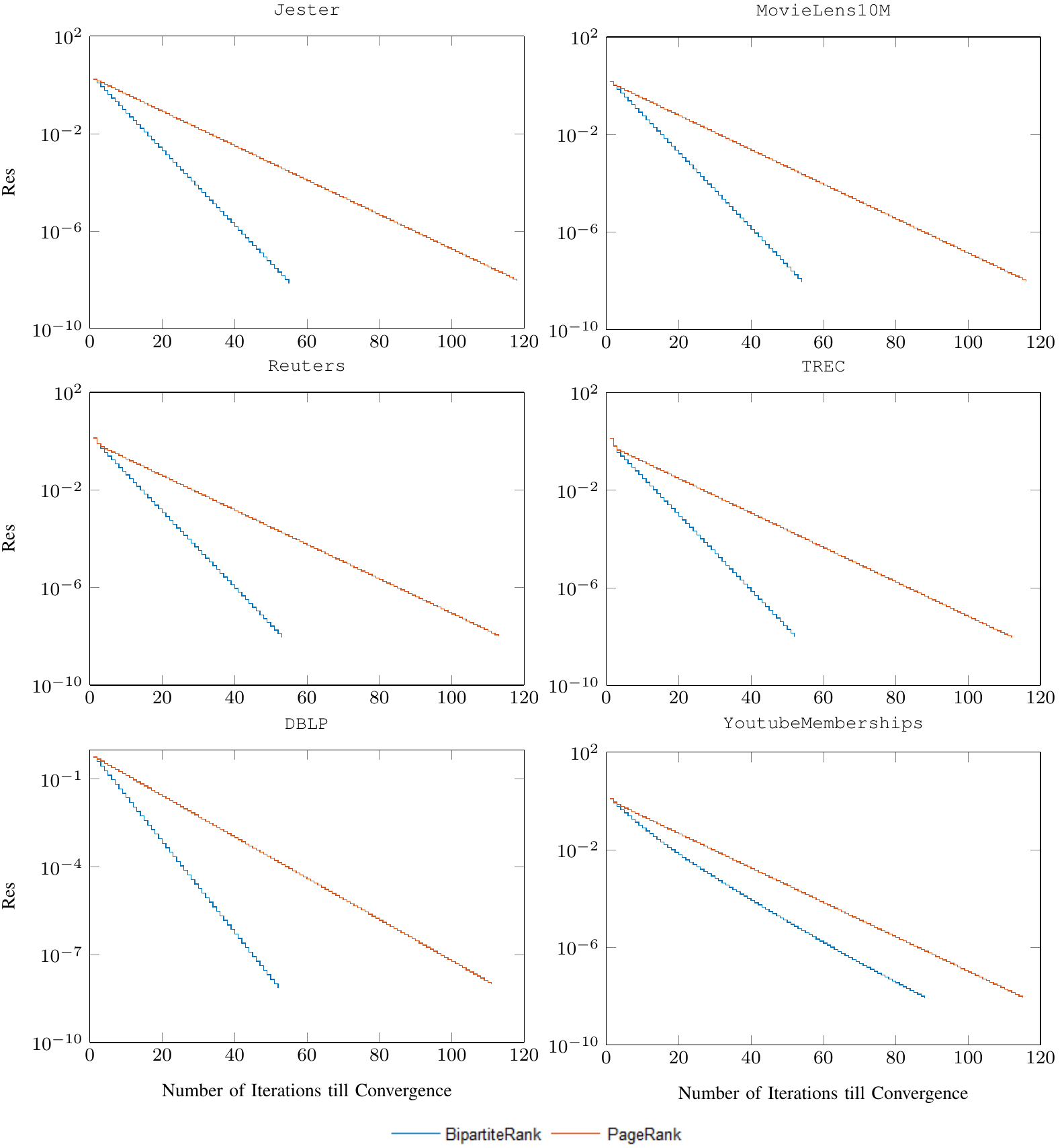} 
\caption{Υπολογιστικά Πειράματα}
\label{fig:computationalTests}
\end{figure}

\begin{figure}
\includegraphics[width=\textwidth]{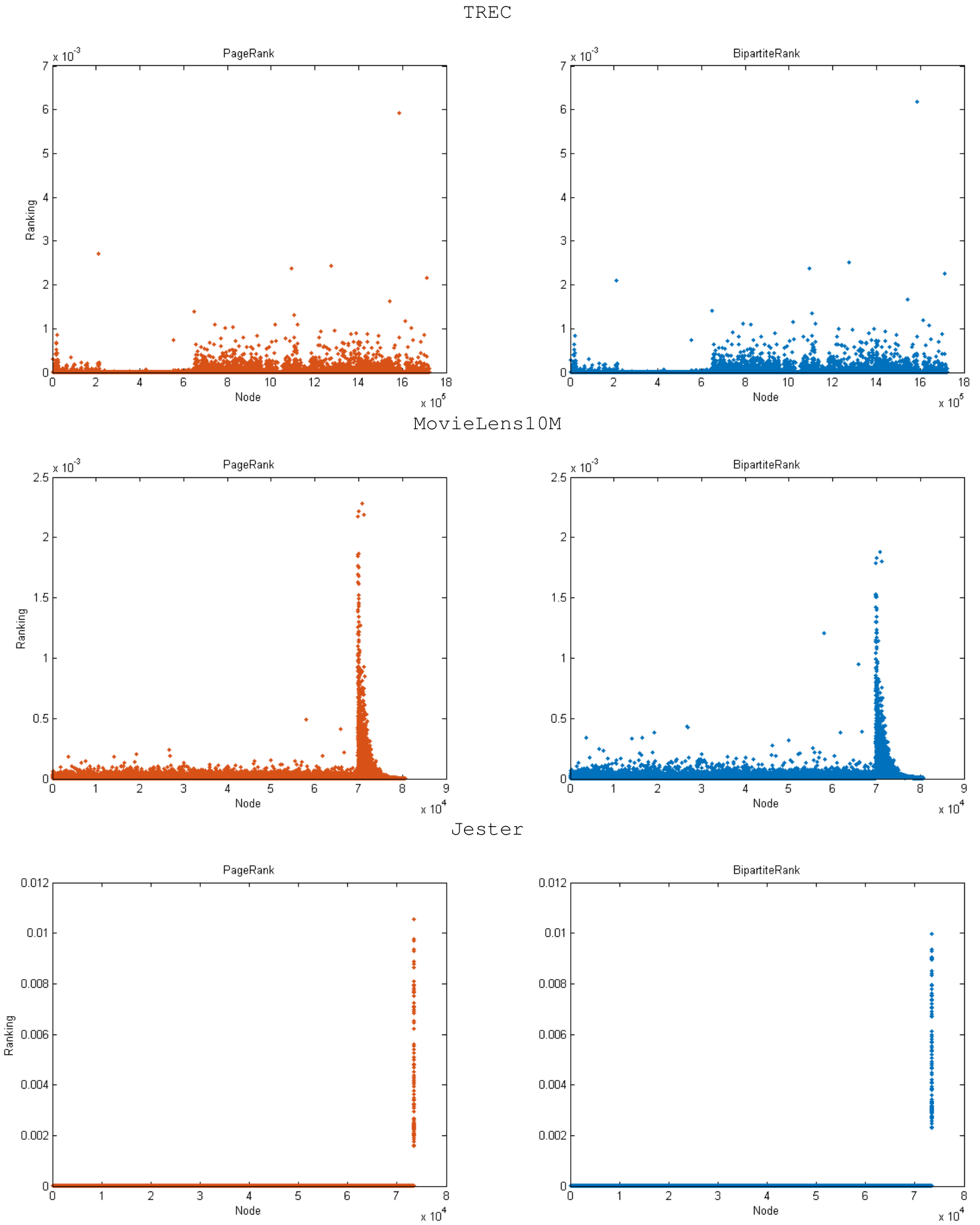}
\caption{Διανύσματα κατάταξης σε τρία σύνολα δεδομένων}
\label{fig:rankingvectors}
\end{figure}

Στο σχήμα $\ref{fig:computationalTests}$ φαίνεται η πορεία των επαναλήψεων μέχρι τη σύγκλιση των αλγορίθμων \en BipartiteRank \gr και \en PageRank\gr, σε όλα τα διαθέσιμα σύνολα δεδομένων, με παράγοντα τηλεμεταφοράς $0.85$. Παρατηρούμε και στην πράξη πως η \en block-wise \gr τηλεμεταφορά  συμβάλλει στην ταχύτερη σύγκλιση (βλέπε ενότητα $\ref{subs:rateofconvergence}$), αφού ο αλγόριθμός μας συγκλίνει πολύ πιο γρήγορα κάνοντας στις περισσότερες περιπτώσεις σχεδόν τις μισές επαναλήψεις σε σχέση με τον \en PageRank\gr.

O παράγοντας τηλεμεταφοράς ελέγχει ουσιαστικά την προτεραιότητα που δίνεται στην δομή υπερσυνδέσμων του γραφήματος, έναντι της τεχνητής τηλεμεταφοράς. Όσον αφορά στον υπολογιστικό του ρόλο, είναι ο παράγοντας που ελέγχει τον ασυμπτωτικό ρυθμό σύγκλισης της δυναμομεθόδου (βλέπε $\ref{subs:rateofconvergence}$ ). Μετρήσαμε τον αριθμό των επαναλήψεων που χρειάζονται μέχρι τη σύγκλιση οι αλγόριθμοι \en BipartiteRank \gr και \en PageRank \gr για διάφορες τιμές του παράγοντα τηλεμεταφοράς, χρησιμοποιώντας όλα τα σύνολα δεδομένων. Τα αποτελέσματα παρουσιάζονται στον πίνακα $\ref{table:varyingparametertests}$, όπου βλέπουμε πως για όλες τις πιθανές τιμές του παράγοντα τηλεμεταφοράς ο αλγόριθμος μας συγκλίνει σε λιγότερες επαναλήψεις.

Τέλος, στο σχήμα $\ref{fig:rankingvectors}$ παρουσιάζονται ενδεικτικά τα διανύσματα κατάταξης των \en TREC, MovieLens10M \gr και \en Jester\gr. Παρατηρούμε κάποιες πολύ μικρές διαφορές μεταξύ των διανυσμάτων κατάταξης των δύο αλγορίθμων, οι οποίες οφείλονται στο γεγονός πως ο αλγόριθμος \en BipartiteRank \gr χειρίζεται την τηλεμεταφορά πιο δίκαια.

%Οι κώδικες που χρησιμοποιήθηκαν για την πραγματοποίηση των πειραμάτων παρατίθενται στο παράρτημα.

\begin{table} 
\centering
\caption{Αριθμός επαναλήψεων μέχρι τη σύγκλιση για διάφορες τιμές του παράγοντα τηλεμεταφοράς}
\en
\scalebox{0.82}{ \begin{tabular}{cccccccccc}
        &\multicolumn{2}{c}{$\eta=0.8$} & \multicolumn{2}{c}{$\eta=0.85$} & \multicolumn{2}{c}{$\eta=0.9$} & \multicolumn{2}{c}{$\eta=0.95$} \\ \hline
Dataset & \textbf{BR}  & PR &  \textbf{BR} & PR &  \textbf{BR} & PR &  \textbf{BR} &  PR \\ \hline
\texttt{MonieLens10M}    & \textbf{38}  &  85 & \textbf{54} & 116  & \textbf{86} & 179  & \textbf{180}  & 367\\ %
\texttt{Jester}   & \textbf{38} &  86 & \textbf{55} & 118  & \textbf{87} & 182  &  \textbf{182} & 373\\ %
\texttt{TREC}     & \textbf{24}  &  51 & \textbf{33} & 69  & \textbf{52} & 107  &  \textbf{108} & 219\\ %
\texttt{Reuters}   & \textbf{38}  &  83 & \textbf{53} & 113 & \textbf{84} & 175 &  \textbf{176} & 358\\ %
\texttt{DBLP}    & \textbf{36}  &  81 & \textbf{52} & 111 & \textbf{82} & 171 &  \textbf{172} & 352\\ %
\texttt{YoutubeMemberships}  & \textbf{66}  &  84 & \textbf{88} & 115 & \textbf{130} & 176 &  \textbf{245} & 362\\ \hline%
\end{tabular}}
\label{table:varyingparametertests}
\end{table}
\gr

\chapter{Συμπεράσματα - Μελλοντική Έρευνα}

Στην παρούσα διπλωματική εργασία, παρουσιάσαμε έναν νέο αλγόριθμο κατάταξης σε διμερή γραφήματα, ο οποίος είναι βασισμένος στον \en PageRank\gr. Αυτό που τον διαφοροποιεί, είναι το γεγονός ότι ξεφεύγει από το κλασικό μοντέλο τηλεμεταφοράς, εισάγοντας ένα άλλο είδος τηλεμεταφοράς που βασίζεται στην \en block \gr δομή του διμερούς γραφήματος. Μετά από πειράματα, αποδείχθηκε πως ο αλγόριθμος μας είναι πολύ πιο αποδοτικός από τον \en PageRank \gr από υπολογιστικής άποψης, αφού συγκλίνει στις περισσότερες περιπτώσεις σχεδόν στα μισά βήματα.

Μια πολύ ενδιαφέρουσα κατεύθυνση που αξίζει να εξερευνηθεί μελλοντικά είναι η σταθερότητα \en (stability) \gr της μεθόδου μας, δηλαδή το πόσο ευαίσθητο είναι το παραγόμενο διάνυσμα κατάταξης σε μικρές αλλαγές στο γράφημα. Επιπλέον, θα μπορούσε να ελεγχθεί η αποδοτικότητα άλλων επαναληπτικών μεθόδων εκτός της δυναμομεθόδου, όπως της μεθόδου \en Jacobi \cite{stewart2009probability}\gr. Τέλος, η μέθοδος που προτείναμε έχει τις προοπτικές να επεκταθεί ώστε να μπορέσει να εφαρμοστεί σε παρόμοια γραφήματα, όπως για παράδειγμα σε γραφήματα με ίδια δομή αλλά με μεγαλύτερο αριθμό \en blocks \gr (κ-μερή γραφήματα).

\addcontentsline{toc}{chapter}{\gr Βιβλιογραφία}
\en

\end{document}